\documentclass[11pt]{article}
\usepackage{amsmath, amssymb, graphics}

\usepackage[nosort]{cite}
\newcommand{\mathsym}[1]{{}}
\usepackage{verbatim}
\usepackage{amsfonts,euscript,amssymb,stmaryrd,braket}
\usepackage{graphics,tikz}
\usetikzlibrary{shapes.misc,arrows,decorations.markings,patterns,calc,shapes.geometric}
\usepackage{caption}
\usepackage{subcaption}
\usepackage{slashed}
\definecolor{hyperref}{RGB}{026,028,185}
\usepackage[bookmarks=true,colorlinks=true,linkcolor=hyperref,citecolor=hyperref,urlcolor=hyperref,bookmarksnumbered]{hyperref}
 \usepackage{booktabs}
 \usepackage{multirow}
\usepackage{tabularx}
\usepackage{longtable}
\usepackage{bbold,bbding}
\usepackage{amsthm} 
\usepackage{esint}
\newcommand{\bal}{\begin{equation}\begin{aligned}}
\newcommand{\eal}{\end{aligned} \end{equation}}
\def\id{\protect{{1 \kern-.28em {\rm l}}}}

\makeatletter
\renewcommand\section{\@startsection {section}{1}{\z@}%
                                   {-3.5ex \@plus -1ex \@minus -.2ex}%
                                   {2.3ex \@plus.2ex}%
                                   {\normalfont\large\bfseries}}
\renewcommand\subsection{\@startsection{subsection}{2}{\z@}%
                                   {-3.25ex\@plus -1ex \@minus -.2ex}%
                                   {1.5ex \@plus .2ex}%
                                   {\normalfont\normalsize\bfseries}}

\makeatother

\numberwithin{equation}{section}

\usepackage[utf8]{inputenc}
\usepackage{epsfig}
\usepackage{graphicx}
\usepackage[multiple]{footmisc}
\usepackage{amssymb,amsmath}
\usepackage{fancybox,framed,tikz}
\usetikzlibrary{decorations.pathmorphing,patterns}
\usepackage{dsfont}
\usepackage{mathtools}
\usepackage{braket}
\usepackage{slashed}
\usepackage{rotating}
\usepackage{bbold,amsfonts}
\usepackage{multirow}
\usepackage{verbatim}
\usepackage{upgreek}

\tikzset{cross/.style={cross out, draw=black, minimum size=2*(#1-\pgflinewidth), inner sep=0pt, outer sep=0pt},
cross/.default={1pt}}

\newcommand{\be}{\begin{equation}}
\newcommand{\ee}{\end{equation}}

\usepackage{color}

\definecolor{mypink1}{rgb}{0.958, 0.188, 0.478}

\newcommand{\ba}{\begin{eqnarray}}
\newcommand{\ea}{\end{eqnarray}}

\hypersetup{}

\tikzset{Witten diagram/.style={execute at begin picture={%
\draw[blue ,fill=blue!05] circle[radius=\pgfkeysvalueof{/tikz/Witten/radius}];
\path node (X){\phantom{X}};
},baseline={(X.base)}},vertex/.style={circle,fill,inner sep=1.414pt,node
contents={}},
Witten/.cd,radius/.initial=1.414cm}

\begin{document}
\renewcommand{\thefootnote}{\arabic{footnote}}

\overfullrule=0pt
\parskip=2pt
\parindent=12pt
\headheight=0in \headsep=0in \topmargin=0in \oddsidemargin=0in

\begin{center}
\vspace{1.2cm}
{\Large\bf \mathversion{bold}
{Analytic bootstrap for the localized magnetic field}\\
}
 
\author{ABC\thanks{XYZ} \and DEF\thanks{UVW} \and GHI\thanks{XYZ}}
 \vspace{0.8cm} {
 Lorenzo~Bianchi$^{a,b}$ \footnote{\tt lorenzo.bianchi@unito.it}, Davide Bonomi$^{c}$ \footnote{{\tt davide.bonomi@city.ac.uk}}, Elia de Sabbata$^{a,b}$ \footnote{{\tt elia.desabbata@unito.it}}
 }
 \vskip  0.5cm

\small
{\em
$^{a}$  
Dipartimento di Fisica, Universit\`a di Torino and INFN - Sezione di Torino\\ Via P. Giuria 1, 10125 Torino, Italy\\    

$^{b}$  
I.N.F.N. - sezione di Torino,\\
Via P. Giuria 1, I-10125 Torino, Italy\\    

$^{c}$   Department of Mathematics, City, University of London,\\
Northampton Square, EC1V 0HB London, United Kingdom
\vskip 0.02cm

}
\normalsize

\end{center}

\vspace{0.3cm}
\begin{abstract} 
We study the two-point function of local operators in the critical $O(N)$ model in the presence of a magnetic field localized on a line. We use a recently developed conformal dispersion relation to compute the correlator at first order in the $\epsilon$-expansion and we extract the full set of defect and bulk CFT data using the Lorentzian inversion formulae. The only input for the computation of the connected correlator is its discontinuity at first order in perturbation theory, which is determined by the anomalous dimension of a single bulk operator. We discuss possible low-spin ambiguities and perform several diagrammatic checks of our results. 
\end{abstract}

\newpage

\tableofcontents
 \newpage  
\section{Introduction and discussion}
The development of conformal bootstrap methods, both numerical and analytical, had an extraordinarily strong impact on the study of statistical models. One of the most exciting progress has certainly been the numerical estimation of the critical exponents of several statistical systems with unprecedented precision \cite{Rattazzi:2008pe,El-Showk:2012cjh,El-Showk:2014dwa,Kos:2016ysd,Poland:2018epd}. In parallel, the discovery of powerful analytical techniques, such as the Lorentzian inversion formula \cite{Caron-Huot:2017vep,Simmons-Duffin:2017nub} and the conformal dispersion relation \cite{Carmi:2019cub}, allowed us to make progress in situations where the conformal field theory contains a small parameter (not necessarily the coupling). With these methods one can obtain results order by order in the expansion parameter only using symmetries and internal consistency. Specifically, for the case of critical models in $(4-\epsilon)$ dimensions, analytic bootstrap techniques can be successfully used to reproduce and extend the results in $\epsilon$-expansion \cite{Alday:2017zzv,Henriksson:2018myn,Alday:2019clp,Carmi:2020ekr,Bertucci:2022ptt}.

While a large wealth of results is now available for correlators of local operators, the study of extended excitations, or defects, with bootstrap techniques has just started to reveal some of its interesting features \cite{Liendo:2012hy,Gliozzi:2015qsa,Liendo:2016ymz,deLeeuw:2017dkd,Rastelli:2017ecj,defectABJM,Drukker:2017dgn,Giombi:2018qox,Bianchi:2018scb,Bianchi:2018zpb,Carlo,Bissi:2018mcq,Kaviraj:2018tfd,Mazac:2018biw,DiPietro:2019hqe,Gimenez-Grau:2019hez,Bianchi:2019dlw,Bianchi:2019sxz,Giombi:2019enr,Wang:2020seq,Bianchi:2020hsz,Ashok:2020ekv,Lauria:2020emq,Behan:2020nsf,Agmon:2020pde,Drukker:2020atp,Herzog:2020bqw,Barrat:2020vch,Dey:2020jlc, Giombi:2021uae,Ferrero:2021bsb,Bianchi:2021snj,Bianchi:2021piu,Cavaglia:2021bnz,Gimenez-Grau:2021wiv,Barrat:2021yvp,Padayasi:2021sik,Behan:2021tcn,Collier:2021ngi,Herzog:2022jqv,Cavaglia:2022qpg,Chalabi:2022qit}. In particular, compared to the homogeneous case, where the crossing equation relates different OPE channels of local operators, the crossing equation for a bulk two-point function in the presence of a defect allows for a cross-talk between two very different OPE expansions \cite{Billo:2016cpy}. In the bulk channel the two operators are expanded in the bulk OPE, i.e. as infinite sum over local bulk operators, while in the defect channel both operators are expanded in terms of an infinite tower of defect operators. This important property has led to the formulation of two different Lorentzian inversion formulae: one that reconstructs the defect spectrum through a discontinuity that is controlled by the bulk OPE \cite{Lemos:2017vnx} and one that does the opposite \cite{Liendo:2019jpu}. Starting from these inversion formulae, whose purpose is to extract the CFT data starting from the (double-)discontinuity of the correlator, one can derive conformal dispersion relations reconstructing the full correlator from its (double-)discontinuity \cite{Barrat:2022psm,Bianchi:2022ppi}.

In this paper we apply these techniques to study the two-point correlator of local operators in the $O(N)$ critical model with a localized magnetic field. The $O(N)$ critical model is a well-known conformal field theory which can be realized by deforming the four-dimensional free theory of $N$ scalar fields by a quartic $O(N)$ invariant interaction tuning the coupling to the Wilson-Fisher fixed point in $(4-\epsilon)$ dimension. Famous applications, i.e. critical models in the same universality class, are the Ising model for $N=1$, the XY model and the Helium superfluid transition for $N=2$ and the isotropic magnets for $N=3$. 

A natural extended excitation in this theory is obtained by switching on a magnetic field along a line, breaking the $O(N)$ symmetry down to $O(N - 1)$ along the defect. This line defect can be nicely realized on the lattice and it has been studied with Monte Carlo simulations in \cite{Allaismagnetic,ParisenToldin2017}.
Also experimental applications are conceivable, either in quantum simulators \cite{Ebadi:2020ldi} or in a mixture of two liquids with a colloidal impurity \cite{ LAW2001159, Fisher2003,ParisenToldin2017}. It is therefore important to produce predictions for the defect CFT data of this critical system. Field-theoretical studies for some observables either at large $N$ or in $\epsilon$-expansion are available in \cite{PhysRevLett.84.2180,Allais:2014fqa,Cuomo:2021kfm}. In particular, the recent study of \cite{Cuomo:2021kfm}, besides providing new results for some observables in the $\epsilon$-expansion, provided a well-defined strategy to study the large $N$ regime finding consistent results in the two expansions. Even more recently \cite{Gimenez-Grau:2022czc} analyzed this defect for $N=3$ using numerical bootstrap techniques for the four-point function of defect operators. 

Here we initiate the analysis of bulk two-point functions, which give access to an infinite set of defect CFT data. The techniques we use seem to rely very little on the specific defect under analysis. Therefore, it is very likely that similar methods could be applied to other line defects in the $O(N)$ critical theory, such as the magnetic impurities considered in \cite{sengupta97,Vojta_2000, Sachdev_2000,Sachdev:2001ky,Sachdev:2003yk,Florens_2006,Florens_2007,Liu_2021} or the twist defects \cite{Gaiotto:2013nva,Billo:2013jda}. Very recently, \cite{Giombi:2022vnz} introduced a class of conformal line defects in fermionic Gross-Neveu-Yukawa models and it would be very interesting to extend our analysis to that case as well.

\subsection*{Summary of the results}
For the $O(N)$ Wilson Fisher theory in $3<d<4$ the fundamental excitation is the vector of scalar fields $\phi_i$ with $i=1,\dots,N$. The main observable in this work is the two-point function
\begin{equation}\label{twoptfuctionointro}
 \langle \phi_i(x) \phi_j(y) \rangle_D ,
\end{equation}
in the presence of the line defect $D$ obtained by coupling the field $\phi_1$ to a magnetic field localized on a line. At the critical point the theory is described by a defect CFT and the bulk correlator \eqref{twoptfuctionointro} is fixed up to two functions of the conformal cross ratios $z$ and $\bar z$ 
 \begin{equation}
  \langle \phi_i(x) \phi_j(y) \rangle_D=\frac{\hat{F}_S(z,\bar{z})\delta_{i1}\delta_{j1}+\hat{F}_V(z,\bar{z})\left(\delta_{ij}-\delta_{i1}\delta_{j1}\right)}{|x_\perp|^{\Delta_\phi}|y_\perp|^{\Delta_\phi}}.
 \end{equation}
In $\epsilon$-expansion, the functions $\hat F_S$ and $\hat F_{V}$ can be expanded perturbatively and one of the goals of this work is to determine them at first order in $\epsilon$. 

Diagrammatically, one can easily see that there is only one connected Feynman diagram contributing to these functions at that order. Therefore, it does not seem to be necessary to appeal to the bootstrap machinery for this computation. Nevertheless, a more careful analysis shows that the diagram is surprisingly hard to compute with ordinary techniques and it does not take values in the space of generalized polylogarithmic functions. Indeed, even its discontinuity is expressed in terms of elliptic functions. From a bootstrap standpoint, this complicated structure has a very clear origin. The non-trivial part of the discontinuity of the functions $\hat F_S$ and $\hat F_{V}$ comes from a single bulk conformal block, associated to the lightest operator exchanged in the bulk channel. The bulk conformal blocks for a line defect in four dimensions are not known in a closed form, but for the specific values $\Delta=2$ and $\ell=0$ that are relevant in this case we can write down the block (and consequently the discontinuity) in terms of incomplete elliptic integrals of the first kind. 

Despite the complicated functional form, the fact that a single bulk operator completely determines the discontinuity is an important conceptual step. Indeed, the dispersion relation developed in \cite{Barrat:2022psm,Bianchi:2022ppi} allows to reconstruct the full correlator from its discontinuity. Since the anomalous dimension of bulk operators are known from the analysis of the homogeneous bulk theory, we can reconstruct the non-trivial part of the one-loop correlator using as our only input a single piece of bulk CFT data. This is a universal feature of defects in the $O(N)$ critical model since it does not require any knowledge of the specifics of the defect. However, the differences among defects may arise in the contributions at low transverse spin. Indeed, the Lorentzian inversion formula may fail to reproduce low spin data if the correlator is not sufficiently well-behaved in a particular limit. Correspondingly, the dispersion relation may lead to a result which misses (possibly infinitely many) low spin contributions. For the case of the localized magnetic field we show by comparison to Feynman diagrams that this low spin problem affects the result very mildly and the only additional piece of CFT data that we need is the one-loop one-point function of the fundamental field $\phi$ which provides a disconnected contribution and was computed in \cite{Allais:2014fqa,Cuomo:2021kfm}. It would be interesting to explore how this low-spin ambiguity is important for other line defects in the $O(N)$ critical model.

The full result for the correlator \eqref{twoptfuctionointro} takes a complicated form, which we spell out in \eqref{correlatormeijer}. Here it is more interesting to show the results for the CFT data of the exchanged operators. In the defect channel, the exchanged operators transform in irreducible representations of the preserved $O(N-1)$ symmetry and in the correlator \eqref{twoptfuctionointro} singlet (S) and vector (V) operators are exchanged. In particular, only the family of operators that already appeared at tree level contribute at one loop because the bulk-to-defect couplings of other operators are suppressed by a power $\epsilon^2$. Therefore, we extracted the one-loop anomalous dimensions $\gamma^{(1)}$ and squared bulk-to-defect couplings $\hat{b}^{2 (1)}$ as a function of the transverse spin $s$.
\begin{align}
         &\hat{\gamma}_{S,0,s}^{(1)}= \frac{1-s}{(2 s+1)}, &
        &\hat{b}_{S,0,s}^{2 (1)} = \frac{-2 (s-1) H_s-3 H_{s+\frac{1}{2}}}{2(2s+1)} \nonumber,\\
        &\hat{\gamma}_{V,0,s}^{(1)}= - \frac{s}{(2 s+1)}, &
        &\hat{b}_{V,0,s}^{2 (1)} = -\frac{(2 s+1) \left(2 s H_s+H_{s-\frac{1}{2}}\right)+2}{2 (2 s+1)^2},
\end{align}
where $H_z$ is the harmonic number.
Notice the presence of zeroes in the anomalous dimensions for $s=1$ in the singlet channel and for $s=0$ in the vector channel. These values correspond to the displacement and the tilt operators respectively. The former is a protected defect operator associated to the explicit breaking of translation invariance, while the latter is a $O(N-1)$ vector associated to the $N-1$ broken generators of the internal symmetry. Therefore, this is an important consistency check of our result.

Knowing the full result one can also extract the defect CFT data appearing in the bulk channel. In that case, the exchanged operators transform in the singlet (S) or in the symmetric traceless (T) representation of $O(N)$. Of course, the anomalous dimensions are already known since they are not affected by the defect. The new data are the one-loop values for the product of the one-point function of the exchanged operator $a_{\mathcal{O}}$ and the bulk OPE coefficients $\lambda_{\phi\phi\mathcal{O}}$. For the twist-two operators that already appeared at tree level we find the correction
\begin{align}
& a \lambda^{(1)}_{T,0,\ell}=N a \lambda^{(1)}_{S,0,\ell}= - \frac{2^{-\ell-7} \Gamma \left(\frac{\ell}{2}+\frac{1}{2}\right)^3(N+8)}{\pi  \Gamma \left(\frac{\ell}{2}+1\right) \Gamma \left(\ell+\frac{1}{2}\right)} \nonumber \times \\ &\times \Big(-32  H_{\frac{\ell}{2}-\frac{1}{2}}+35  H_{\ell-\frac{1}{2}}+19  \psi ^{(0)}(\ell) -38  \psi ^{(0)}(2 \ell)-19 \gamma  +38 \log (2)+16 \frac{N^2-3N-22}{(N+8)^2}\Big),
\end{align}
where $\psi^{(0)}(z)$ is the digamma function. 
In principle, also twist-four operators can be exchanged. Nevertheless, while at twist two there is only a single family of operators in the OPE of two fundamental fields $\phi_i$, starting from twist four one can create more than one primary for a given number of fields and derivatives. These operators would enter the OPE with their classical dimensions and therefore they are degenerate at this order in perturbation theory. To lift the degeneracy one would need to compute additional correlators. Still, one can extract the quantities $a \lambda^{(1)}_{S,1,\ell}$ and $a \lambda^{(1)}_{T,1,\ell}$ (our results are given in \eqref{bulkdata1}), which correspond, for the degenerate cases, to linear combinations of the defect CFT data associated to the single degenerate operators. Furthermore, since the bulk OPE coefficients are already at order $\epsilon$, the only new information is about the tree-level one-point functions of twist-four operators, which can be easily obtained also by other means. 

The plan of the paper is as follows: in Section \ref{sec:review} we review the $O(N)$ critical model in the presence of a localized magnetic field, in Section \ref{sec:bootstrap} we briefly present the dispersion relation and Lorentzian inversion formulae, which we use in Section \ref{sec:treelevel} and \ref{sec:oneloop} to compute the correlator and extract the defect CFT data at tree level and at one loop, respectively. Some technical details and diagrammatic checks are contained in the Appendix.

{\bf Note added:} While this paper was in preparation, we became aware of \cite{toappear}, whose content partially overlaps with the present work. We coordinated with the author for a simultaneous submission.

\section{The O(N) critical model with a localized magnetic field}
\label{sec:review}

The starting point for the construction of the $O(N)$ critical model is the Euclidean action of $N$ free massless scalar fields in the fundamental representation of $O(N)$ perturbed by a $O(N)$-invariant quartic interaction
\begin{equation}
S=\int d^d x \left[ \frac{1}{2}\left( \partial_\mu \phi_i \right)^2+\frac{\lambda_0}{4!}\left(\phi_i \phi_i\right)^2 \right],
\end{equation}
where $d=4-\epsilon$. Since $d<4$ the perturbation is relevant and it triggers a renormalization group flow. This flow can be studied by working pertirbatibvely in $\lambda_0$; in particular we adopt the minimal subtraction (MS) scheme. The beta function of the renormalized coupling constant $\lambda$ at one loop reads
\begin{equation}
\beta(\lambda)=\frac{\partial \lambda}
{\partial \log \mu}=-\epsilon \lambda +\frac{N+8}{48\pi^2}\lambda^2+O(\lambda^3),
\end{equation}
where $\mu$ is the mass scale introduced in the renormalization process. If the mass term is fine-tuned to zero, this flow admits the well known infrared Wilson-Fisher fixed point \cite{Wilson:1971dc}, which describes a conformal field theory. This fixed point corresponds to the non-trivial zero of the beta function
\begin{equation}
\beta(\lambda_*)=0 \quad \Rightarrow \quad \lambda_* =\frac{48 \pi^2}{N+8}\epsilon + \mathcal{O}(\epsilon^2).
\end{equation}
In the perturbative setup, the operators of this conformal field theory are just the renormalized version of the local operators constructed using the bare fields. For example, the renormalized field $[\phi_i](x)$ is given by
\begin{equation}
[\phi_i](x)=Z_\phi \,\phi_i (x),
\end{equation}
where in this case the wavefunction renormalization $Z_\phi$ is just a factor since there is no operator mixing. The wavefunction renormalization can be computed by imposing the finiteness of the correlators. In particular, at one loop it is readily found that
\begin{equation}
Z_\phi(\lambda)=1+O(\lambda^2).
\end{equation}
From the wavefunction renormalization it is possible to extract the anomalous dimension of the field $\phi_i$ in the following way
\begin{equation}
\gamma_\phi(\lambda)=-\frac{\partial \log Z_\phi}{\partial \log \mu}=-\beta(\lambda)\frac{\partial \log Z_\phi}{\partial \lambda}= 0 + O(\lambda^2).
\end{equation}
The conformal dimension $\Delta_\phi$ of the operator $[\phi_i]$ is \cite{Henriksson:2022rnm}
\begin{equation}
\Delta_\phi=\frac{d-2}{2}+\gamma_\phi(\lambda_*)=1-\frac{\epsilon}{2}+\mathcal{O}(\epsilon^2).
\end{equation}
Note that at this order, the conformal dimension coincides exactly with the engineering dimension of the bare field, since the anomalous dimension vanishes. We can now consider the case in which we deform the action also by an additional perturbation localized on a line
\begin{equation}
S_D=h_0 \int d\tau \, |\dot{x} (\tau)| \phi_1 (x(\tau)),
\end{equation}
where $x(\tau)$ describes a line as the real parameter $\tau$ varies, which we denote by $D$ (called defect), and $h_0$ is a new coupling constant. In particular, we will consider the case in which $D$ is a straight line (or equivalently, a circle).
This perturbation explicitly breaks the $O(N)$ global symmetry of the model down to $O(N-1)$. For the free theory in four dimensions, this perturbation produces a simple example of a conformal defect as the operator $\phi$ has dimension one and $h_0$ is a defect marginal parameter \cite{Kapustin:2005py}. As we move away from four dimension, the bulk theory flows to the Wilson-Fisher fixed point and the operator $\phi$ is a weakly relevant defect deformation (the bulk dimension is $4-\epsilon$, but the defect dimension is fixed to one). Therefore, this perturbation together with the quartic interaction triggers a renormalization group flow in the two coupling constants. One can study this joint flow using standard diagrammatic techniques without the need to work perturbatively in $h_0$, since any diagram contributing to any correlator at some fixed order in $\lambda$ will contain insertion of $h_0$ only up to a finite power. That is to say, $h_0$ is not considered to be a small coupling constant. This flow admits an infrared fixed point for the following value\footnote{
As one can find in \cite{Cuomo:2021kfm}, to calculate the value of $h_*$ to the first order in $\epsilon$ one need to consider perturbations up to two loops. However, for the purpose of this work the small parameter entering in the analytic bootstrap is $\epsilon$, hence we do not truncate results at one or two loops but rather at order one in $\epsilon$.
} 
of the renormalized coupling constants \cite{Allais:2014fqa,Cuomo:2021kfm}
\begin{equation}
\lambda_* =\frac{48 \pi^2}{N+8}\epsilon + \mathcal{O}(\epsilon^2), \quad h_*=\sqrt{N+8}+\frac{4N^2+45N+170}{4(N+8)^\frac{3}{2}}\epsilon+\mathcal{O}(\epsilon^2).
\end{equation}
The defect perturbation breaks explicitly the conformal symmetry group $SO(d+1,1)$ of the system at the fixed point. Since we have chosen the defect $D$ to be a straight line, the system will still be invariant under a residual symmetry group $SO(2,1)\times SO(d-1)$, generated by conformal transformation on the line and rotation around the defect $D$. This is equivalent to saying that the fixed point under consideration can be described by a (line) defect conformal field theory.

\subsection{The observable: bulk two-point function}

The main observables of this defect conformal field theory (dCFT) are correlators of local operators in the presence of the defect. From this section onwards every local operator will be assumed to be already renormalized. There are two kind of local operators: bulk operators and defect operators. Bulk operators are well-defined in those points of the space which do not lie in the defect $D$, whereas defect operators have $D$ as their support. In the $\epsilon$-expansion, both kinds of operators arise as the renormalized version of composite operators evaluated at the fixed point. The residual conformal symmetry group $SO(2,1)\times SO(d-1)$ imposes severe constraints on the correlators through Ward identities analogous to the homogeneous case. The constraints are of course weaker due to the presence of the defect $D$. For instance, one-point functions of generic bulk operators $O_{\Delta, \mu_1 \dots \mu_\ell }(x)$ do not vanish, but instead they take the form \cite{Billo:2016cpy}
\begin{equation}\label{oneptfunction}
\langle \mathcal{O}_{\Delta, \mu_1 \dots \mu_\ell }(x) \rangle_D = \frac{a_\mathcal{O}}{|x_\perp|^\Delta}I_{\mu_1 \dots \mu_\ell}(x_\perp),
\end{equation}
where $x_\perp$ is the projection of $x$ in the subspace orthogonal to the defect $D$, $\Delta$ is the conformal dimension of the operator and $I_{\mu_1 \dots \mu_\ell}(x_\perp)$ is a completely determined tensor\footnote{
The specific form of this tensor $I_{\mu_1 \dots \mu_l}(x_\perp)$ can be easily determined using the embedding formalism, see e.g. \cite{Billo:2016cpy}.
}.
The constants $a_\mathcal{O}$ are real and they are part of the data of the dCFT. In our model, there is an additional global $O(N-1)$ symmetry which imposes further constraints on the correlators. For example, the one-point function of $\phi_i(x)$ must be proportional to the only $O(N-1)$ invariant tensor with one free index, namely $\delta_{i1}$,
\begin{equation}
\langle \phi_i(x) \rangle_D = \delta_{i1} \frac{a_{\phi}}{|x_{\perp}|^{\Delta_{\phi}}}.
\end{equation}
A much more interesting correlator, on which this work will focus, is the bulk two-point function of two fundamental fields
\begin{equation}
\langle \phi_i(x) \phi_j(y) \rangle_D, \quad \quad x,y \notin D.
\end{equation}
Conformal and global symmetries constrain this correlator to take the following form
 \begin{equation}\label{twoptfuction}
 \langle \phi_i(x) \phi_j(y) \rangle_D=\frac{F_1(z,\bar{z})\delta_{ij}+F_2(z,\bar{z})\delta_{i1}\delta_{j1}}{|x_\perp|^{\Delta_\phi}|y_\perp|^{\Delta_\phi}},
 \end{equation}
where $F_1(z,\bar{z})$ and $F_2(z,\bar{z})$ are arbitrary functions of the two conformally invariant cross-ratios $z$ and $\bar z$ defined by
\begin{equation}
\frac{(x-y)^2}{|x_\perp||y_\perp|}=\frac{(1-z)(1-\bar{z})}{\sqrt{z\bar{z}}}, \quad \frac{x\cdot 
 y}{|x_\perp||y_\perp|}= \frac{z+\bar{z}}{2\sqrt{z\bar{z}}}.
\end{equation}
The geometrical interpretation of $z$ and $\bar z$ becomes transparent if we use conformal transformations to put the two bulk points on a plane orthogonal to the defect and then to move one of the two to $(1,0)$ on this plane. Then, in Euclidean signature, $z$ and $\bar z$ are complex coordinates for the second point on this plane. It is also useful to introduce radial coordinates $r$ and $w$ on the orthogonal plane
\begin{equation}
 \quad z=rw, \quad \bar{z}=\frac{r}{w}.
\end{equation}
Rotating to Lorentzian signature with time orthogonal to the defect, $z$ and $\bar z$ become real and independent lightcone coordinates. 

It is well known that in a CFT there exists an operator product expansion (OPE) for any couple of primary operators. This OPE is convergent inside correlation functions provided that there are no other operator insertions closer to the point of expansion than the two primaries. In a dCFT, the same OPE of the bulk theory holds with the additional condition that the defect also has to be sufficiently distant from the point of expansion. In addition to the bulk OPE however, there is also a new channel, called defect channel \cite{Billo:2016cpy}. It consists of an expansion of a bulk operator as an infinite sum of defect operators, and it is convergent inside correlation functions whenever the operator is close enough to the defect. In general, these two expansions can be further refined to manifestly account for the structure of the internal symmetry of the theory. 

For the bulk channel, the OPE of two operators transforming under the representations $R_1$ and $R_2$ of the internal symmetry group must contain exchanged operators in an irreducible representation contained in the Clebsch-Gordan decomposition of $R_1\otimes R_2$. In the case of our model, for two fundamental scalar fields it is possible to write symbolically
 \begin{equation}\label{ope}
 \phi_i(x)\phi_j(y)=\sum_{\Delta, \ell , R}  \, \lambda_{\phi \phi \mathcal{O}}^{ij \, a} \, |x-y|^{\Delta-2\Delta_\phi} \left( \mathcal{O}_{\Delta, \ell}^{\,a}(y) + \text{descendants} \right),
 \end{equation}
 where $\Delta_\phi$ is the conformal dimension of the field $\phi$. The conformal dimension and the spin of the exchanged primary operator $\mathcal{O}_{\Delta, \ell}^{\,a}$ are labelled by $\Delta$ and $\ell$ respectively, and $R$ denotes the $O(N)$ irreducible representations contained in the tensor product $V\otimes V$ of two vector representations V, i.e. the singlet $S$, the symmetric traceless $T$ and the antisymmetric $A$. The index $a$ runs over $a= 1,\dots , \text{dim}R$. In \eqref{ope} we have suppressed spacetime indices. Since this is a OPE between scalar operators, only operators in even spin-$\ell$ traceless symmetric representations of $SO(d)$ can appear in the decomposition. The tensor structure of the three-point function coefficients $\lambda_{\phi \phi \mathcal{O}_S}^{ij}$ can be easily written down for $R=S,T,A$ as 
 \begin{equation}\label{OPEcoefficientstructure}
 \begin{split}
 \lambda_{\phi \phi \mathcal{O}_S}^{ij}= \lambda_{\phi \phi \mathcal{O}_S}\delta_{ij}, \quad \lambda_{\phi \phi \mathcal{O}_T}^{ij \, (kl)}=\lambda_{\phi \phi \mathcal{O}_T}\left(\delta_{i(k}\delta_{jl)}-\frac{1}{N}\delta_{ij}\delta_{kl}\right), \quad \lambda_{\phi \phi \mathcal{O}_A}^{ij \, [kl]}=\lambda_{\phi \phi \mathcal{O}_A}\delta_{i[k}\delta_{jl]},
 \end{split}
 \end{equation}
 where the (anti-)symmetrization has been taken with weight $1/2$.
 
 On the other hand, the operators appearing in the defect OPE will be organized into $O(N-1)$ representations since the $O(N)$ symmetry is explicitly broken by the defect. In general, if the symmetry group $G$ is broken down to a subgroup $H$ by the defect and if a bulk operator sits in the representation $R$ of $G$, the operators exchanged in the defect OPE must transform in an irreducible representations contained in the branching rule of the restricted representation $R^{(G)}\rightarrow R^{(H)}$. Hence we can write
 \begin{equation}
 \phi_i(x)=\sum_{\hat{\Delta},s,R} b^{\, i \, a}_{\phi\hat{\mathcal{O}}} |x_\perp|^{\hat{\Delta}-\Delta_\phi}\left(\hat{\mathcal{O}}^{\,a}_{\hat{\Delta},s}(x)+ \text{descendants} \right),
 \end{equation}
 where $\hat{\Delta}$ and $s$ are the conformal dimension and the transverse spin of the defect operator $\hat{\mathcal{O}}_{\hat{\Delta}, s}^{\, a}$ respectively. Since $\phi_i(x)$ is a scalar, the operators exchanged in this OPE do not carry longitudinal spin. From the branching rule $V^{O(N)}\rightarrow S^{O(N-1)}\oplus V^{O(N-1)}$ one can immediately see that the only allowed representations $R$ are the singlet $S$ and the vector $V$ of $O(N-1)$. For these two representations, the tensor structures of the bulk-defect two-point function coefficients $b^{\, i \, a}_{\phi\hat{\mathcal{O}}}$ are
\begin{equation}\label{bcoeffstructure}
b^{\, i }_{\phi\hat{\mathcal{O}}_S}=b_{S,\hat{\Delta},s}\delta_{i1}, \quad \quad b^{\, i \, \hat{j}}_{\phi\hat{\mathcal{O}}_V}=b_{V,\hat{\Delta},s}\delta_{i\hat{j}},
\end{equation}
where $\hat{j}=2,\dots, N$, and $\delta_{i1}$ and $\delta_{i\hat{j}}$ are projectors from the representation space of $V^{O(N)}$ to those of $S^{O(N-1)}$ and $V^{O(N-1)}$ respectively.
 
These two different OPE decompositions lead to two conformal block expansions, which we analyze in the following section.

\subsection{Block expansions}
To analyze the bulk block expansion, it is convenient to rewrite the bulk two-point function as
\begin{equation}\label{FSFT}
 \langle \phi_i(x) \phi_j(y) \rangle_D=\frac{F_S(z,\bar{z})\delta_{ij}+F_T(z,\bar{z})\left(\delta_{i1}\delta_{j1}-\frac{1}{N}\delta_{ij}\right)}{|x_\perp|^{\Delta_\phi}|y_\perp|^{\Delta_\phi}},
\end{equation}
 where $F_S(z,\bar{z})$ and $F_T(z,\bar{z})$ are linear combinations of the functions $F_1(z,\bar{z})$ and $F_2(z,\bar{z})$ introduced in \eqref{twoptfuction} (see \eqref{linearcombinations}).
 In terms of lightcone coordinates, this bulk channel decomposition reads
 \begin{equation}\label{bulkchannelexpansion}
F_S(z,\bar{z})\delta_{ij}+F_T(z,\bar{z})\left(\delta_{i1}\delta_{j1}-\frac{1}{N}\delta_{ij}\right)=\left(\frac{\sqrt{z\bar{z}}}{(1-z)(1-\bar{z})}\right)^{\Delta_\phi}\!\!\!\sum\limits_{\substack{\Delta, \ell \\ R=S,T}}  \lambda_{\phi \phi \mathcal{O}}^{ij \, a} \, a_{\mathcal{O}}^{\,a} \, f_{\Delta,\ell}(z,\bar{z}),
 \end{equation}
 where the explicit form of the bulk conformal blocks $f_{\Delta,\ell}(z,\bar{z})$ was found in \cite{Isachenkov:2018pef} and it is given in Appendix \ref{app:kinematics}. Note that the only bulk operators which can have a non vanishing one-point function are those for which the identity appears as an exchanged operator in their defect OPE. Therefore, the allowed tensor structures for the coefficients $a_\mathcal{O}^{\,a}$ are obtained by projecting $O(N)$ representations into the singlet $S^{O(N-1)}$. It immediately follows that operators in antisymmetric representations have zero one-point functions. For this reason, the sum in \eqref{bulkchannelexpansion} is taken only over the singlet $S$ and the symmetric traceless $T$ representations of $O(N)$. The tensor structures of the one-point functions for $R=T,S$ are
  \begin{equation}\label{Oneptfunctionstructure}
 \begin{split}
 a_{ \mathcal{O}_S}, \quad  a_{ \mathcal{O}_T}^{(ij)}=a_{ \mathcal{O}_T}\left(\delta_{i1}\delta_{j1}-\frac{1}{N}\delta_{ij}\right).
 \end{split}
 \end{equation}
 Inserting \eqref{OPEcoefficientstructure} and \eqref{Oneptfunctionstructure} into \eqref{bulkchannelexpansion} one gets the following block decompositions
 \begin{equation}\label{bulkblockexpansion}
 \begin{split}
 F_S(z,\bar{z})=\left(\frac{\sqrt{z\bar{z}}}{(1-z)(1-\bar{z})}\right)^{\Delta_\phi}\sum_{\Delta,  \ell }  \lambda_{\phi\phi\mathcal{O}_S}a_{\mathcal{O}_S} f_{\Delta,\ell}(z,\bar{z}),\\
  F_T(z,\bar{z})=\left(\frac{\sqrt{z\bar{z}}}{(1-z)(1-\bar{z})}\right)^{\Delta_\phi}\sum_{\Delta,  \ell }  \lambda_{\phi\phi\mathcal{O}_T}a_{\mathcal{O}_T} f_{\Delta,\ell}(z,\bar{z}).
 \end{split}
 \end{equation}
 
 In a similar fashion, for the defect channel it is helpful to rewrite the bulk two-point function in the following way
 \begin{equation}\label{FSFV}
  \langle \phi_i(x) \phi_j(y) \rangle_D=\frac{\hat{F}_S(z,\bar{z})\delta_{i1}\delta_{j1}+\hat{F}_V(z,\bar{z})\left(\delta_{ij}-\delta_{i1}\delta_{j1}\right)}{|x_\perp|^{\Delta_\phi}|y_\perp|^{\Delta_\phi}},
 \end{equation}
  where again $\hat{F}_S(z,\bar{z})$ and $\hat{F}_V(z,\bar{z})$ are linear combinations of $F_1(z,\bar{z})$ and $F_2(z,\bar{z})$ given in \eqref{linearcombinations}. The defect channel decomposition is 
  \begin{equation}
   \hat{F}_S(z,\bar{z})\delta_{i1}\delta_{j1}+\hat{F}_V(z,\bar{z})\left(\delta_{ij}-\delta_{i1}\delta_{j1}\right)=\sum\limits_{\substack{\hat{\Delta}, s \\ R=S,V}}   b^{\, i \, a}_{\phi\hat{\mathcal{O}}} b^{\, j \, a}_{\phi\hat{\mathcal{O}}} \hat{f}_{\hat{\Delta},s},
  \end{equation}
where the explicit form of the defect conformal blocks $\hat{f}_{\hat{\Delta},s}(z,\bar{z})$ is given in Appendix \ref{app:kinematics}. Using \eqref{bcoeffstructure} one gets
 \begin{equation}\label{defectblockexpansion}
 \begin{split}
\hat{F}_S(z,\bar{z})=\sum_{\hat{\Delta},  s }  b_{S,\hat{\Delta},s}^2 \hat{f}_{\hat{\Delta},s},\\
\hat{F}_V(z,\bar{z})=\sum_{\hat{\Delta},  s }  b_{V,\hat{\Delta},s}^2\hat{f}_{\hat{\Delta},s}.
 \end{split}
 \end{equation}
 
 Everything said so far holds at the non-perturbative level. However, more can be said if one looks at the interplays between the block expansions \eqref{bulkblockexpansion} and \eqref{defectblockexpansion} with the perturbative series. For instance, consider the three-point function coefficients $\lambda_{\phi\phi\mathcal{O}}$. If the operator $\mathcal{O}$ is composed with an even\footnote{In the odd $2k+1$ case the coefficient simply vanishes for the representation theory reasons stated in the previous subsection.} number $2(k+1)$ of fundamental fields, then a straightforward diagrammatic argument immediately implies that this coefficient is at least of order $k$ in $\epsilon$. In particular, to the first order in $\epsilon$ only operators with $k=0$ or $k=1$ will enter in the bulk block expansion of the functions $F_S(z,\bar{z})$ and $F_T(z,\bar{z})$. Moreover, only the anomalous dimensions of operators with $k=0$ are relevant at this order, since only the classical dimensions of operators with $k=1$ will contribute to the expansions.
 
  In the next sections it will be shown that from the analytic properties of the correlator \eqref{twoptfuction} it is possible to extract an infinite set of data about this dCFT. In particular, we will be able to extract the product of one-point function coefficients $a_\mathcal{O}$ and bulk three-point functions to the first order in $\epsilon$ for an infinite family of operators, called twist-two operators. The twist of a local operator $O_{\Delta, \mu_1 \dots \mu_\ell }(x)$ with dimension $\Delta$ and spin $\ell$ is defined as $\tau:=\Delta-\ell$.  Twist-two operators (operators with $\tau=2+O(\epsilon)$) are exactly the operators composed using two fundamental fields that we mentioned above. When $\ell \geq 1$ they can be thought as the renormalized version of the infinite weakly broken higher spin currents which are conserved in the UV free theory. Their explicit expression has been found in \cite{Giombi:2016hkj} and it is given by
 \begin{equation}\label{doubletwistexplicit}
 \begin{split}
 J^{a}_{\mu_1 \dots \mu_\ell}(x)&=\mathcal{N}_\ell^R\, P^{\,a}_{ij}  \,\sum_{n=0}^\ell c_{\ell\, n} \,  \partial_{\{ \mu_1} \dots \partial_{\mu_{\ell-n}} \phi_i\, \partial_{\mu_{\ell-n+1}} \dots \partial_{\mu_\ell \} } \phi_j (x), \\
 c_{\ell \,n}&=\frac{(-1)^n}{n! (\ell -n)! \Gamma(n+\frac{d}{2}-1)\Gamma (\ell -n +\frac{d}{2}-1)},
  \end{split}
 \end{equation}
 where brackets denote traceless symmetrization and $P^{\,a}_{ij}$ is a projector into an irreducible representation $R$ of $O(N)$ labelled by the index $a=1, \dots \dim R$; since it is made out of just two indices it can only be $S$, $T$ or $A$. Moreover, from the form of the coefficients $c_{\ell \, n}$ one can easily check that for even (odd) $\ell$ the operator $J^{a}_{\mu_1 \dots \mu_\ell}(x)$ is (anti-)symmetric in the indices $i$ and $j$, hence it is non vanishing only for the representations $S$ and $T$ ($A$). $\mathcal{N}_\ell^R$ is a factor which ensures that the two-point function of $J^{a}_{\mu_1 \dots \mu_\ell}(x)$ is correctly normalized. The linear combination in \eqref{doubletwistexplicit} make sure that the operators are in a diagonal basis with respect to the dilatation operator, hence they are primary operators of the dCFT. In the spin $\ell=0$ case, there are two operators
 \begin{equation}
 \phi^2(x) = \mathcal{N}_0^S \phi_i \phi_i(x), \quad T_{ij}(x)=\frac{1}{2}\mathcal{N}_0^T \left(\phi_i \phi_j -\frac{\delta_{ij}}{N}\phi_k \phi_k \right) (x).
 \end{equation}
Their anomalous dimensions already  appear at first order in $\epsilon$ and they are
\begin{equation}
\gamma_{S,0} = \frac{N+2}{N+8} \epsilon + O(\epsilon^2), \quad \gamma_{T,0} = \frac{2}{N+8}\epsilon+O(\epsilon^2).
\end{equation}
On the other hand, in the spin $\ell \geq 1$ case one can show (see e.g. \cite{Giombi:2016hkj}) that all the anomalous dimensions vanish at first order in $\epsilon$, and that they are non zero only from the second order onwards 
\begin{equation}
\gamma_{R,\ell} = 0+  O(\epsilon^2), \quad \forall \,  \ell \geq 1.
\end{equation}
This will be the crucial point of this work since the bulk anomalous dimensions are precisely the data we need to compute the discontinuity of the correlator and to reconstruct it through the dispersion relation, which we now review.

\section{Lorentzian inversion formulae and dispersion relation}\label{sec:bootstrap}
A powerful way to study homogeneous CFTs analytically is the Lorentzian inversion formula derived in \cite{Caron-Huot:2017vep}. It is an integral formula which reconstructs the OPE data of any CFT from the double discontinuity of four-point functions. In favorable situations, the double discontinuity receives contributions only from very few operators and the inversion formula can extract from it an infinite amount of CFT data \cite{Alday:2017vkk,Alday:2019clp} .\\
For the defect case, there are two analogous formulae. A defect inversion formula was derived in \cite{Lemos:2017vnx} and allows to extract the defect channel CFT data from a single discontinuity
\begin{equation}\label{defectinversion}
    \begin{split}
         b(\hat{\Delta},s) = \int_{0}^{1}  \frac{dz}{2z} z^{-\frac{\hat{\tau}}{2}} \int_{1}^{\frac{1}{z}}  \frac{d \bar{z}}{2 \pi i} &(1-z \bar{z}) (\bar{z}-z) \bar{z}^{- \frac{\hat{\Delta}+s}{2}-2} \ {}_2 F_1 \left(s+1, 2- \frac{q}{2},\frac{q}{2}+s, \frac{z}{\bar{z}}\right) \times\\
         & \times {}_2 F_1 \left(1-\hat{\Delta}, 1-\frac{p}{2},1+\frac{p}{2}-\hat{\Delta}, z \bar{z}\right) \text{Disc}\,F(z,\bar{z}).
    \end{split}
\end{equation}
The coefficient function $b(\hat{\Delta},s)$ has simple poles for $\hat{\Delta}$ equal to the dimensions of exchanged operators and residues given by the defect OPE coefficients $b^2_{\hat{\Delta},s}$. Therefore $b(\hat{\Delta},s)$ contains all the CFT data of the exchanged defect operators. The crucial ingredient of this inversion formula is the discontinuity
\begin{equation}\label{discontinuitydef}
 \text{Disc} F(z,\bar{z})=F(z,\bar{z}+i \epsilon)-F(z,\bar{z}-i \epsilon),
\end{equation}
where $F(z,\bar{z}+i \epsilon)$ and $F(z,\bar{z}-i \epsilon)$ indicate that $\bar{z}$ should be taken above or below the branch cut at $\bar z=1$, leaving $z$ fixed.
The defect inversion formula was derived in \cite{Lemos:2017vnx} through a contour deformation argument, which is justified only if the integrand vanishes sufficiently fast for large $w$, or equivalently for $w\to 0$ since the correlator is symmetric under $w \leftrightarrow \frac{1}{w}$. More precisely, this implies that the formula \eqref{defectinversion} is valid for transverse spin $s>s_*$ if
\begin{equation}\label{smallw}
    F(r,w)\sim w^{-s_*},  \quad \quad  w\to 0.
\end{equation}
This means that in general the inversion formula may miss contributions to the CFT data from low spin operators.

There is also a bulk inversion formula, which allows to extract the bulk OPE data from a double discontinuity~\cite{Liendo:2019jpu}. The formula reads \footnote{Compared to \cite{Liendo:2019jpu}, we rewrite the bulk inversion formula for $\left(\frac{(1-z)(1-\bar{z})}{\sqrt{z \bar{z}}}\right)^{\Delta_\phi} F(z,\bar z)$ instead of $F(z,\bar z)$ in order to use a more conventional definition of double discontinuity.}
\begin{equation} \label{bulkinversion}
\begin{split}
     &c(\Delta,\ell)=  c^{t}(\Delta,\ell)+(-1)^\ell  c^{u}(\Delta,\ell), \\
     &c^{t}(\Delta,\ell)=\frac{\kappa_{\Delta+\ell}}{2}\int_0^1 d^2 z \ \mu(z,\bar z)\  f_{\ell+d-1,\Delta-d+1}(z,\bar z) \text{dDisc}\left( \left(\frac{(1-z)(1-\bar{z})}{\sqrt{z \bar{z}}}\right)^{\Delta_\phi}F(z,\bar z) \right),
\end{split}
\end{equation}
with
\begin{equation}
\begin{split}
    &\kappa_{\Delta+\ell}=\frac{\Gamma(\frac{\Delta+\ell}{2})^4}{2\pi^2\Gamma(\Delta+\ell)\Gamma(\Delta+\ell-1)}, \\
    &\mu(z,\bar z)= \frac{|z-\bar z|^{d-p-2}|1-z \bar z|^p}{(1-z)^d (1-\bar z)^d},
\end{split}
\end{equation}
and where $f_{\Delta,\ell}(z,\bar z)$ are the bulk blocks. The $u$-channel term is the same with the two external bulk operators exchanged. 
In this case, the coefficient function $c(\Delta,\ell)$ has poles corresponding to the dimensions of the operators that are exchanged in the bulk OPE and corresponding residues given by the product of bulk three-point functions and one-point functions, $\lambda_{\phi\phi \mathcal{O}}a_{\mathcal{O}}$. The input of the formula is the double discontinuity defined by
\begin{equation}\label{ddiscdef}
     \text{dDisc}F(z,\bar z)=F(z,\bar z)-\frac12 F^{\circlearrowleft}(z,\bar z)-\frac12 F^{\circlearrowright}(z,\bar z),
\end{equation}
where this time the functions $F^{\circlearrowleft}(z,\bar z)$ and $F^{\circlearrowright}(z,\bar z)$ are obtained by taking the analytic continuation around the point $\bar z=0$, leaving $z$ fixed. Just like the defect inversion formula, the bulk inversion formula might fail for low spins \cite{Liendo:2019jpu}. More precisely, the formula is valid for spins $\ell>\ell_{*}$ where
\begin{equation}\label{smallwl}
    \left(\frac{(w-r)(1-w r)}{r w}\right)^{\Delta_{\phi}}F(r,w) \lesssim w^{1-\ell_{*}}, \quad \quad w \rightarrow 0.
\end{equation}
The two Lorentzian inversion formulae allow to extract the defect CFT data of the theory from certain discontinuities of the two-point functions. Therefore, these discontinuities contain all the information that is necessary to reconstruct the full correlator. This is made explicit in the dispersion relation, a formula that computes the full correlator directly from a discontinuity \cite{Carmi:2019cub}. In the case of defect CFTs, the dispersion relation reads ~\cite{Bianchi:2022ppi,Barrat:2022psm}
\begin{equation}\label{dispersion}
     F(r,w)=\int_{0}^{r} \frac{d w'}{2\pi i}\left(\frac{1}{w'-w}+\frac{1}{w'-\frac{1}{w}}-\frac{1}{w'}\right) \text{Disc}F(r,w'),
\end{equation}
where $\text{Disc }F(r,w)$ is the discontinuity through the cut running from $w=0$ to $w=r$. From the definition of the variable $w$, we see that $\text{Disc}F(r,w)=-\text{Disc}F(z,\bar{z})$ as defined in \eqref{discontinuitydef}. Just like the defect inversion formula  \eqref{defectinversion}, this dispersion relation is derived from Cauchy's theorem by deforming the contour around the singularities and dropping the contributions at infinity. This contour deformation argument  misses terms that are given by low spin conformal blocks, which give contributions at infinity. If one knows the behaviour of the correlator for $w\to 0$ (or equivalently for $|w|\to \infty$) \eqref{smallw}, one can take into account these terms by introducing a prefactor in front of the correlator
\begin{equation}
    \tilde{F}(r,w) = \left(\frac{r}{(w-r)(\frac{1}{w}-r)}\right)^{s_*+1} F(r,w).
\end{equation}
By construction, $\tilde{F}(r,w)$ goes like $w^{-1}$ at large $w$ and therefore we can safely ignore the contribution at infinity and reconstruct this rescaled correlator from \eqref{dispersion}. From the point of view of the original correlator, this implies an improved dispersion relation
\begin{equation}\label{improveddisp}
 \frac{F(r,w)}{(w-r)^{s_*+1}(\frac{1}{w}-r)^{s_*+1}}=\int_{0}^{r} \frac{d w'}{2\pi i}\left(\frac{1}{w'-w}+\frac{1}{w'-\frac{1}{w}}-\frac{1}{w'}\right) \text{Disc}\left[\frac{F(r,w')}{(w'-r)^{s_*+1}(\frac{1}{w'}-r)^{s_*+1}}\right].
\end{equation}
In the following, we would like to apply the formulae introduced in this section to the $\epsilon-$expansion of correlators in the critical $O(N)$ model. We start by introducing the leading-order correlator.

\section{Tree level}\label{sec:treelevel}
At leading order in the $\epsilon$-expansion, the correlator contains two terms: the free correlator without the defect which contributes to $F_1$ in \eqref{twoptfuction} and the square of the one-point function \eqref{oneptfunction} which gives the leading contribution to $F_2$ \footnote{The defect coupling at the fixed point is not small so the one-point function of local operators are no suppressed in the $\epsilon$-expansion.}. In particular, the perturbative expansion of $a_{\phi}$ in \eqref{oneptfunction} reads \cite{Allais:2014fqa,Cuomo:2021kfm}
\begin{equation}\label{1ptphi}
    a^{2}_{\phi} = \frac{N+8}{4} + \epsilon \frac{  \left(N^2-3 N+(N+8)^2 \log (4)-22\right)}{8 (N+8)} + \mathcal{O}(\epsilon^2),
\end{equation}
and the leading-order contribution to the two-point function is
\begin{align}\label{treelevelF1F2}
 F_1^{(0)}(r,w)&=\frac{r w}{(r-w) (r w-1)}, \quad \quad F_2^{(0)}(r,w)={a^{2}_{\phi}}^{(0)}.
\end{align}
Each of these two terms has a simple interpretation in one of the two channels. The free correlator $F_1^{(0)}$ corresponds to the exchange of the identity operator in the bulk channel, while the squared one-point function is associated to the exchange of the defect identity in the defect channel. On the other hand, as usual, to reproduce the identity in a given channel, an infinite tower of operators is needed in the crossed one. Let us review the CFT data of these exchanged operators.

Using the linear combinations in \eqref{linearcombinations} one can rewrite \eqref{treelevelF1F2} in terms of $F_S$ and $F_T$ introduced in \eqref{FSFT}
\begin{equation}\label{F0bulk}
\begin{split}
     &F^{(0)}_S(r,w)= \frac{{a^{2}_{\phi}}^{(0)}}{N}+\frac{r w}{(r-w) (r w-1)}, \\
     &F^{(0)}_T(r,w)= {a^{2}_{\phi}}^{(0)}.
\end{split}     
\end{equation}
From this expression we notice the obvious fact that the bulk identity contributes only to the singlet exchange. On the other hand, the constant term ${a^{2}_{\phi}}^{(0)}$ can be reproduced by the exchange of two infinite towers of twist-two spin-$\ell$ operators of the schematic form \footnote{By this symbolic notation we only want to indicate the number of derivatives associated to a given twist-two primary operator.}
\begin{align}
    [\phi^2]_{S,0,\ell}= \phi^i \partial_{\mu_1}... \partial_{\mu_\ell} \phi_i, \quad \quad [\phi^i\phi^j]_{T,0,\ell}= \phi^{(i} \partial_{\mu_1}... \partial_{\mu_\ell} \phi^{j)}-\textup{trace}.
\end{align} 
Their CFT data can be extracted simply by comparing \eqref{F0bulk} with the block expansion \eqref{bulkblockexpansion} or from the Lorentzian inversion formula \eqref{bulkinversion} using \footnote{As we mentioned, the Lorentzian inversion formula does not work for low spins and in this case it does not reproduce the contribution of the bulk identity. Nevertheless, it still reproduces correctly all the defect CFT data of the twist-two operators.} \\ 
\begin{equation}\label{disc0}
    \begin{split}
      &\text{dDisc}\left(\left(\frac{(1-z)(1-\bar{z})}{\sqrt{z \bar{z}}}\right)^{\Delta_\phi} F^{(0)}_S(z,\bar{z})\right) = 2 \frac{{a^{2}_{\phi}}^{(0)}}{N} \frac{(1-z)(1-\bar{z})}{\sqrt{z \bar{z}}},\\
      &\text{dDisc}\left(\left(\frac{(1-z)(1-\bar{z})}{\sqrt{z \bar{z}}}\right)^{\Delta_\phi} F^{(0)}_T (z,\bar{z})\right) = 2 {a^{2}_{\phi}}^{(0)} \frac{(1-z)(1-\bar{z})}{\sqrt{z \bar{z}}}.
    \end{split}
\end{equation}

 The result is~\footnote{The twist-two operators have the same dimensions in both channel at tree level, however they are distinct operators and have different anomalous dimensions and OPE coefficients.} \cite{Liendo:2019jpu}
\begin{equation}\label{bulkdata0}
\begin{split}
    &\Delta^{(0)}_{S,0,\ell} =  \Delta^{(0)}_{T,0,\ell}= 2 \Delta_\phi +\ell, \\
     &a \lambda^{(0)}_{T, k,\ell}= a \lambda^{(0)}_{S,k,\ell}\ N= \delta_{k,0}  \ {a^{2}_{\phi}}^{(0)} \frac{ 2^{-\ell} \Gamma \left(\frac{\ell+1}{2}\right)^3}{\pi  \Gamma \left(\frac{\ell}{2}+1\right) \Gamma \left(\ell+\frac{1}{2}\right)}.
\end{split}
\end{equation}
where the spin $\ell$ is even. Let us comment on our labels for the operators, which will remain the same also at order $\epsilon$. We use the index $k$ to label the classical twist $\tau^{(0)}=\Delta^{(0)}-\ell$, specifically $k=\frac{\tau^{(0)}}{2}-\Delta_{\phi}$.
In particular, at this order, only two families of operators enter in the bulk OPE: the identity and the twist-two operators ($k=0$). Furthermore, for the specific case $k=0$ we know that there is a single primary operator for a given spin $\ell$. This is no longer true for higher values of $k$, where degeneracies may appear. Therefore, our notation will not be able to distinguish among degenerate operator, but since also our observable is not able to make this distinction we find this notation convenient.


Moving on to the defect channel, we have 
\begin{equation}\label{F0}
\begin{split}
    &\hat{F}^{(0)}_{S}(r,w) = {a^{2}_{\phi}}^{(0)} +\frac{r w}{(r-w) (r w-1)}, \\
    &\hat{F}^{(0)}_{V}(r,w) = \frac{r w}{(r-w) (r w-1)},
\end{split}
\end{equation}
where we notice again that the defect identity only enters the singlet channel, while the bulk identity is reproduced by two infinite towers of defect operators of the schematic form
\begin{align}
 [\phi]_{S,0,s}=\partial_{\perp}^s \hat{\phi}, \quad \quad  [\phi^{\hat{i}}]_{V,0,s}&=\partial_{\perp}^s \hat{\phi}^{\hat{i}}.
\end{align}
These operators have transverse spin $s$ and transverse twist $\hat{\tau}= \hat{\Delta}-s=1$. As before one can compare \eqref{F0} with the block expansion \eqref{defectblockexpansion} or use the inversion formula \eqref{defectinversion} with \footnote{Also in this case the Lorentzian inversion formula fails to reproduce the contribution of the identity.}
\begin{equation}
    \begin{split}
    &\text{Disc}\hat{F}^{(0)}_S(r,w) = 2\pi i \Big(\frac{r w}{1- r w}\Big) \delta(r-w), \\
     &\text{Disc}\hat{F}^{(0)}_V(r,w) = 2\pi i \Big(\frac{r w}{1- r w}\Big) \delta(r-w). \\
    \end{split}
\end{equation}
The result is \cite{Lemos:2017vnx, Billo:2016cpy}
\begin{equation}\label{defectdata0}
\begin{split}
    &\hat{\Delta}_{S,0,s}^{(0)} =\hat{\Delta}_{V,0,s}^{(0)}= 1+s, \\
    &\hat{b}^{2  (0)}_{S,m,s}=\hat{b}^{2  (0)}_{V,m,s} =  \delta_{m,0}.
\end{split}
\end{equation}
As in the bulk case we use an additional label $m$ for the classical transverse twist, $m=\frac{\hat{\tau}^{(0)}-1}{2}$. In this case, only twist-one operators ($m=0$) appear in the OPE, as expected from the equation of motion of the bulk field \cite{Lemos:2017vnx}. 

The aim of this work is to find the next-to-leading order corrections to these results and this is the topic of the next section.

\section{One loop}\label{sec:oneloop}
The idea of the defect analytic bootstrap is to use information from the bulk theory to compute correlators in the presence of the defect. In particular, the discontinuity relevant to the dispersion relation \eqref{dispersion} is controlled by the bulk block expansion, hence we analyse which operators appear in the perturbative expansion of \eqref{bulkblockexpansion}. We consider the following perturbative expansion of the bulk channel CFT data
\begin{equation}\label{perturbativedatabulk}
\begin{split}
        \Delta &= \Delta^{(0)}+\epsilon\, \gamma^{(1)}+\mathcal{O}(\epsilon^2), \\
         \lambda_{\phi\phi \Delta} a_{\Delta} &=  a \lambda^{(0)} +\epsilon\,  a \lambda^{(1)} + \mathcal{O}(\epsilon^2). \\
\end{split}        
\end{equation}
More precisely, operators that appeared at tree level, namely twist-two operators, enter in the one-loop OPE with their anomalous dimensions 
\begin{equation}
    \Delta_{S/T,0,\ell} = 2\Delta_\phi + \ell+\epsilon \gamma^{(1)}_{S/T,0,\ell} + \mathcal{O}(\epsilon^2),
\end{equation}
which are known from previous work on the $O(N)$ model without defects (see for example \cite{Henriksson:2022rnm} and references therein)
\begin{align} \label{nondefectdata}
     &\gamma^{(1)}_{S,0,\ell} = \frac{N+2}{N+8} \delta_{0,\ell},&
     &\gamma^{(1)}_{T,0,\ell} = \frac{2}{N+8} \delta_{0,\ell}.
\end{align}
Notice the crucial fact that only the spin-one operator of each representation has a non-vanishing anomalous dimension at one loop. This will be extremely important for the computation of the discontinuity.

The second contribution we need to consider originates from the anomalous dimension of the external operator $\phi_i$
\begin{equation}
    \Delta_\phi = 1-\frac{\epsilon}{2}+ \epsilon \gamma^{(1)}_\phi + \mathcal{O}(\epsilon^2),
\end{equation}
which gives a correction to the bulk identity contribution
\begin{equation}\label{bulkidentity}
\begin{split}
   \textstyle  \left(\frac{r w}{(r-w) (r w-1)}\right)^{\Delta_\phi} =\frac{r w}{(r-w) (r w-1)}+\epsilon \left(\gamma^{(1)}_{\phi}-\frac{1}{2}\right)\frac{r w}{(1-r w)(w-r)} \log \Big( \frac{r w}{(1-r w)(w-r)}\Big)+\mathcal{O}(\epsilon^2).
\end{split}     
\end{equation}
Specifically, the anomalous dimension $\gamma^{(1)}_{\phi}$ is well-known to be vanishing at one loop, i.e. $\gamma^{(1)}_{\phi}= 0$, but clearly this does not kill the contribution $\eqref{bulkidentity}$.

In principle, we can also have operators with higher twist, which appear in the OPE with their classical dimensions
\begin{equation}
\begin{split}
     \Delta_{S/T,k,\ell} = 2\Delta_\phi + 2k+\ell+ \mathcal{O}(\epsilon), \ \ \ \ k>0.
\end{split}
\end{equation}
In particular, we expect only operators of twist up to $4$, since the bulk OPE coefficients of higher-twist operators are of order $\epsilon^2$ \cite{Alday:2017zzv,Henriksson:2018myn}. Nevertheless, while there is a single family of operators with twist two, i.e. a single primary operator for each spin, starting from twist four there can be degenaracies which cannot be lifted by studying a single correlator. Luckily, these operators do not contribute to the discontinuity in \eqref{dispersion} so this is not an issue to reconstruct the full correlator. We will comment further on this issue in Section \ref{bulkdata}.

All in all, the one-loop bulk block expansion for the singlet reads
\begin{equation}\label{bulkopepertS}
\begin{split}
   F^{(1)}_S(r,w) &=  -\frac{r w}{2(1-r w)(w-r)} \log \Big( \frac{r w}{(1-r w)(w-r)}\Big)+ \\ &+ \frac{r w}{2(1-r w)} a \lambda^{(0)}_{S,0,0} \gamma^{(1)}_{S,0,0}   \left(\tilde f_{2,0}(r,w) \log(w-r)+\partial_{\Delta} \tilde f_{2,0}(r,w)\right)+ \\ & +  \sum_{k=0}^{1} \sum_{\ell} \  \frac{r w}{1-r w} a \lambda^{(1)}_{S,k,\ell} (w-r)^k \tilde f_{2+2k+\ell,\ell}(r,w),
\end{split}
\end{equation}
where we defined $ \tilde f_{\Delta,\ell}(r,w)=(w-r)^{\frac{-\Delta+\ell}{2}}f_{\Delta,\ell}$, with $f_{\Delta,\ell}$ given in \eqref{bulkblock}. For the symmetric traceless part we have 
\begin{equation}\label{bulkopepertT}
\begin{split}
   F^{(1)}_T(r,w) &=\frac{r w}{2(1-r w)} a \lambda^{(0)}_{T,0,0} \gamma^{(1)}_{T,0,0}  \left(\tilde f_{2,0}(r,w) \log(w-r)+ \partial_{\Delta} \tilde f_{2,0}(r,w)\right)+ \\ & +  \sum_{k=0}^{1} \sum_{\ell} \  \frac{r w}{1-r w} a \lambda^{(1)}_{T,k,\ell} (w-r)^{k} \tilde f_{2+2k+\ell,\ell}(r,w).
\end{split}
\end{equation}
We can consider an analogous expansion for the defect channel CFT data
\begin{equation}\label{perturbativedatadefect}
\begin{split}
        \hat{\Delta} &= \hat{\Delta}^{(0)}+\epsilon\, \hat{\gamma}^{(1)}+\mathcal{O}(\epsilon^2), \\
         \hat{b} &=  \hat{b}^{(0)} +\epsilon\,  \hat{b}^{(1)} + \mathcal{O}(\epsilon^2).
\end{split}        
\end{equation}
As before, operators with transverse twist one, which already appeared at tree level, enter the block expansion with their anomalous dimension
\begin{equation}
    \hat{\Delta}_{S/V,0,s} = 1+s+\epsilon\, \hat{\gamma}^{(1)}_{S/T,0,s}+\mathcal{O}(\epsilon^2),
\end{equation}
while higher-twist operators contribute with their classical dimensions
\begin{equation}
\begin{split}
     \hat{\Delta}_{S/V,m,s} = 1+2m+s +\mathcal{O}(\epsilon),  \ \ \ \ m>0.
\end{split}
\end{equation}
In particular, at one loop we expect only operators of transverse twist one, i.e. $m=0$, since higher-twist defect operators have bulk-to-defect couplings $\hat{b}$ of order $\epsilon$, hence their squared coefficients are at least of order $\epsilon^2$. We will see later on that this expectation is confirmed by the inversion formula.
The defect block expansions at one loop read
\begin{equation}\label{defectopepert}
\begin{split}
    &\hat{F}^{(1)}_S(r,w) = { a^{2}_{\phi}}^{(1)}+ \sum_s  \ \hat{b}^{2  (1)}_{S,0,s} \hat{f}_{1+s,s}(r,w)+  \hat{b}^{2  (0)}_{S,0,s} \hat{\gamma}^{(1)}_{S,0,s} \partial_{\hat{\Delta}} \hat{f}_{1+s,s}(r,w),  \\
   &\hat{F}^{(1)}_V(r,w) =   \sum_s  \ \hat{b}^{2  (1)}_{V,0,s} \hat{f}_{1+s,s}(r,w)+\hat{b}^{2  (0)}_{V,0,s} \hat{\gamma}^{(1)}_{V,0,s} \partial_{\hat{\Delta}} \hat{f}_{1+s,s}(r,w).
\end{split}
\end{equation}
Before showing results for the defect CFT data let us compute the full one-loop correlator using the dispersion relation.

\subsection{The full result}
We can use the dispersion relation \eqref{dispersion} to compute the full correlator at one loop provided the correlator is sufficiently well-behaved for $w\to 0$ as discussed in \eqref{smallw}. In particular, we assume that after subtracting the defect identity (which is just the one-loop correction to the squared one-point function $a_{\phi}^2$), the rest of the correlator vanishes for $w\to 0$, i.e. that  $s_* < 0$ in \eqref{smallw}. We checked that our assumption is correct by comparing the non-trivial part of the full correlator obtained from the dispersion relation \eqref{correlatormeijer} with the numerical integration of the result from Feynman diagrams in Appendix \ref{app:feynman}. We found perfect agreement between the two results.

The main ingredient of the dispersion relation is the discontinuity \eqref{discontinuitydef}. The discontinuity can be computed term by term in the bulk-channel block expansion. It is convenient to consider the singlet and traceless contribution separately. Starting from the singlet representation, we see from \eqref{bulkopepertS} that the discontinuity  at first order in the $\epsilon$-expansion is given only by the logarithmic terms, because the blocks $\tilde f_{\Delta,\ell}(r,w)$ and their derivatives, evaluated at the tree-level dimensions \eqref{bulkdata0}, are regular at $w=r$. Crucially, these terms are proportional to tree-level data and one-loop anomalous dimensions. Notice that \eqref{nondefectdata}, combined with \eqref{bulkopepertS}, implies that the one-loop discontinuity of the singlet is just given by two terms: one is the correction to the bulk identity from the engineering dimension of the external field and the other one is proportional to a single bulk block.
Using 
\begin{equation}
    \text{Disc}(\log(w-r)) = 2 \pi i,
\end{equation}
and the explicit form of the bulk blocks as a series expansion \eqref{bulkblock} we find
\begin{equation}\label{discellipticS}
\begin{split}
    &\text{Disc} F^{(1)}_S(r,w)=-\pi i \frac{(r w)}{ (r-w) (r w-1)}+2 \pi i \Big(\frac{r w}{2(1-r w)}\Big) a \lambda^{(0)}_{S,0,0} \gamma^{(1)}_{S,0,0}  \tilde f_{2,0}(r,w)= \\
     &= -\textstyle \pi i \frac{(r w)}{ (r-w) (r w-1)} + \frac{N+2}{8 N} \frac{4 \pi  \sqrt{r w} \left(F\left(\sin ^{-1}\left(\sqrt{r} \sqrt{\frac{r-w}{1-r w}}\right)|\frac{(r w-1)^2}{(r-w)^2}\right)-F\left(\sin ^{-1}\left(\frac{\sqrt{\frac{r-w}{1-r w}}}{\sqrt{r}}\right)|\frac{(r w-1)^2}{(r-w)^2}\right)\right)}{r-w},
\end{split} 
\end{equation}
where $F(x,k)$ is the incomplete elliptic integral of the first kind. We stress that this contribution is given by a single operator in the bulk expansion, i.e. the operator of twist two and spin zero $\phi^2$, whose anomalous dimension \eqref{nondefectdata} is our only input. The complicated form of the function \eqref{discellipticS} is due to the bulk conformal block $\tilde f_{2,0}$ which is a particular case of the complicated expression in \eqref{bulkblock}.

The first term in \eqref{discellipticS} trivially reproduces the correction to the bulk identity  \eqref{bulkidentity}, as we expected. The second contribution is very complicated and we could not solve the corresponding integral in the dispersion relation in terms of simple functions. However, we can find the result as a series expansion or as a derivative of a special function. We refer to \eqref{app:explicit} for the explicit computation. At the end of the day we find
\begin{equation}\label{correlatormeijer}
    \begin{split}
       & \textstyle F^{(1)}(r,w)_{S,\text{not id}} = \frac{N+2}{8 N} \sum_{m=0}^{\infty}  \frac{2^{1-m} r w (1-r^2)^m G_{4,4}^{4,2}\left(\frac{4 w}{w r^2-\left(w^2+1\right) r+w}|
\begin{array}{c}
 0,0,\frac{m}{2},\frac{m+1}{2} \\
 -\frac{1}{2},0,m,m \\
\end{array}
\right)}{(m!)\left(r^2 (-w)+r w^2+r-w\right)} =\\
&= \textstyle \frac{N+2}{8 N} \frac{\partial }{\partial t} \left( \left(\frac{(r-w) (r w-1)}{(1+r)^2 w}\right)^t \frac{4 r}{(1+r)^2 (2 t+1)}  \left. F^{112}_{ 101}\left(
\begin{array}{c}
 1+t:\frac{1}{2};\frac{1}{2}+t,1; \\
 \frac{3}{2}+t:-;1+t; \\
\end{array}
\ \left(\frac{1-r}{1+r}\right)^2,\frac{(r-w) (r w-1)}{(1+r)^2 w}\right) \right) \right|_{t=0}\\
& \equiv  \frac{N+2}{8 N} I(r,w),
    \end{split}
\end{equation}
where the result is expressed in terms of an infinite sum of Meijer G-functions $G_{4,4}^{4,2}$ or a derivative of a Kampè de Fèriet function $F_{101}^{112}$ \cite{KdFconvergence}, whose definitions are given in Appendix \ref{app:explicit}. In the last line of \eqref{correlatormeijer} we defined the function $I(r,w)$ for later convenience.
Putting all together and adding back the contribution of the defect identity, the singlet contribution reads
\begin{equation}\label{F1S}
\begin{split}
     F^{(1)}_S(r,w) &= \frac{{a^{2}_{\phi}}^{(1)}}{N}-\Big(\frac{r w}{2(1-r w)(w-r)}\Big) \log \Big( \frac{r w}{(1-r w)(w-r)}\Big) + \frac{N+2}{8 N} I(r,w).
\end{split}
\end{equation}
We can follow the same argument for the symmetric traceless part, the only difference is that the bulk identity contribution is absent and the overall factor changes because of the OPE data \eqref{bulkdata0} and \eqref{nondefectdata}. The discontinuity reads
\begin{equation}\label{discellipticT}
\begin{split}
    &\text{Disc} F^{(1)}_T(r,w)=2 \pi i \Big(\frac{r w}{2(1-r w)}\Big) a \lambda^{(0)}_{T,0,0} \gamma^{(1)}_{T,0,0}  \tilde f_{2,0}(r,w)= \\
     &=  \frac{\pi  \sqrt{r w} \left(F\left(\sin ^{-1}\left(\sqrt{r} \sqrt{\frac{r-w}{1-r w}}\right)|\frac{(r w-1)^2}{(r-w)^2}\right)-F\left(\sin ^{-1}\left(\frac{\sqrt{\frac{r-w}{1-r w}}}{\sqrt{r}}\right)|\frac{(r w-1)^2}{(r-w)^2}\right)\right)}{r-w}.
\end{split} 
\end{equation}
At the end of the day we find
\begin{equation}\label{F1T}
\begin{split}
     F^{(1)}_T(r,w) &= {a^{2}_{\phi}}^{(1)} + \frac{1}{4} I(r,w).
\end{split}
\end{equation}
Using the combinations in \eqref{linearcombinations} we can rewrite the results in terms of the functions $\hat{F}_S$ and $\hat{F}_T$ that are natural in the defect channel as
\begin{equation}
\begin{split}
    & \hat{F}_{S}^{(1)}(r,w) = {a^{2}_{\phi}}^{(1)}-\Big(\frac{r w}{2(1-r w)(w-r)}\Big) \log \Big( \frac{r w}{(1-r w)(w-r)}\Big) + \frac{3}{8} I(r,w),\\
     & \hat{F}_{V}^{(1)}(r,w) = -\Big(\frac{r w}{2(1-r w)(w-r)}\Big) \log \Big( \frac{r w}{(1-r w)(w-r)}\Big) + \frac{1}{8} I(r,w).
\end{split}
\end{equation}
We now move to the computation of the defect CFT data which appear in the two channels.

\subsection{Defect channel data}
In principle, we could extract the defect CFT data in the defect channel by plugging the discontinuity in the defect inversion formula \eqref{defectinversion}, under the same assumptions as in the case of the dispersion relation, namely that we are not missing any low spin contribution except for the defect identity. However, as usual, solving the integral to find the full coefficient function $b(\hat{\Delta},s)$ is very hard. For this reason, we follow the approach of \cite{Caron-Huot:2017vep}, which was specialized to the case of defects in \cite{Lemos:2017vnx}, and perform a small $z$ expansion in the inversion formula and only then we compute the integral for each term in the expansion.
If we expand the defect inversion integral at small $z$ we find
\begin{equation}\label{defectinversionseries}
\begin{split}
     &b(\hat{\Delta},s) = \int_{0}^{1}  \frac{dz}{2z} z^{-\frac{\hat{\tau}}{2}} \sum_{m=0} z^{m} \sum_{k=-m}^{m} c_{m,k}(\hat{\Delta},s) B(z,\beta+2k), \\
     & B(z,\beta) = \int_{1}^{\infty}  \frac{d \bar{z}}{2 \pi i} \bar{z}^{-\frac{\beta}{2}-1} \text{ Disc}\hat{F}_{S,V}(z,\bar{z}),
\end{split}
\end{equation}
where $\beta = \hat{\Delta}+s = \hat{\tau} +2s$ and where $c_{m,k}(\hat{\Delta},s)$  are the coefficients obtained from the small $z$ expansion of the integrand of \eqref{defectinversion}. Notice that a term proportional to $z^{\alpha}$ in the small $z$ expansion reproduces the contribution of twist $2 \alpha$ in the coefficient function $b(\hat{\Delta},s)$, because the last integral is
\begin{equation}
    \int_0^1 dz \ \frac{z^{-\frac{\hat{\tau} }{2}}}{2 z} z^{\alpha } = -\frac{1}{\hat{\tau} -2 \alpha }.
\end{equation}
We will show explicitly the computation of the coefficient only for the leading order in the small $z$ expansion, which allows to extract the CFT data for transverse twist equal to $1$. We will just give the results for higher orders because the expressions become more complicated. \\
Starting from \eqref{discellipticS} and \eqref{discellipticT} and performing the linear combinations \eqref{linearcombinations}, we obtain 
\begin{equation}\label{disc1}
\begin{split}
    & \textstyle \text{Disc}\hat{F}_{S}^{(1)}(r,w) = - \frac{\pi i(r w)}{ (r-w) (r w-1)} +  \frac{3 \pi  \sqrt{r w} \left(F\left(\sin ^{-1}\left(\sqrt{r} \sqrt{\frac{r-w}{1-r w}}\right)|\frac{(r w-1)^2}{(r-w)^2}\right)-F\left(\sin ^{-1}\left(\frac{\sqrt{\frac{r-w}{1-r w}}}{\sqrt{r}}\right)|\frac{(r w-1)^2}{(r-w)^2}\right)\right)}{2(r-w)}, \\
     & \textstyle \text{Disc}\hat{F}_{V}^{(1)}(r,w) = -  \frac{\pi i (r w)}{ (r-w) (r w-1)} +  \frac{ \pi  \sqrt{r w} \left(F\left(\sin ^{-1}\left(\sqrt{r} \sqrt{\frac{r-w}{1-r w}}\right)|\frac{(r w-1)^2}{(r-w)^2}\right)-F\left(\sin ^{-1}\left(\frac{\sqrt{\frac{r-w}{1-r w}}}{\sqrt{r}}\right)|\frac{(r w-1)^2}{(r-w)^2}\right)\right)}{2(r-w)}.
\end{split}
\end{equation}
We see that in both the singlet and the vector defect representations we have two contributions in the discontinuity  \footnote{Notice that in both channels the discontinuity  does not depend on $N$, therefore the only data in the defect OPE that will depend on $N$ will be the defect identity correction, which is missed by the inversion formula.}. The first one comes from the the bulk identity operator. This contribution was already considered in \cite{Lemos:2017vnx} for a generic value of $\Delta_{\phi}$ and it gives
\begin{equation}
    \begin{split}
        B (z,\beta)_{\text{id}} = \frac{\sin (\pi  \Delta_{\phi} ) \Gamma (1-\Delta_{\phi} ) \left(-\frac{\sqrt{z}}{z-1}\right)^{\Delta_{\phi} } \Gamma \left(\frac{\beta +\Delta_{\phi} }{2}\right)}{\pi  \Gamma \left(\frac{1}{2} (\beta -\Delta_{\phi} +2)\right)}.
    \end{split}
\end{equation}
Expanding in $\epsilon$ and selecting the first order gives
\begin{equation}\label{Bid}
    \begin{split}
        B(z,\beta)^{(1)}_{\text{id}} &= \frac{\sqrt{z} \left(\psi ^{(0)}\left(\frac{\beta +1}{2}\right)+\log \left(-\frac{\sqrt{z}}{z-1}\right)+\gamma \right)}{2 (z-1)}=\\
        &= \frac{1}{4} \sqrt{z} \left(-2 \psi ^{(0)}\left(\frac{1+\beta}{2}\right)-\log (z)-2 \gamma \right) +O(z^{\frac{3}{2}}),
    \end{split}
\end{equation}
where $\psi ^{(0)} (z)$ is the digamma function.
This result has to be combined with the second contribution in equation \eqref{disc1}, which can be expanded as
\begin{equation}\label{disczseries}
\begin{split}
     \text{Disc}\hat{F}_S (z,\bar{z})_{\text{not id}}= \frac{3}{4} i \pi  \sqrt{z} \left(\log (z)+\log (\bar{z})-4 \log \left(\frac{2 \sqrt{\bar{z}}}{\sqrt{\bar{z}}+1}\right)\right)+ O(z^{\frac{3}{2}}), \\
     \text{Disc}\hat{F}_V (z,\bar{z})_{\text{not id}}= \frac{1}{4} i \pi  \sqrt{z} \left(\log (z)+\log (\bar{z})-4 \log \left(\frac{2 \sqrt{\bar{z}}}{\sqrt{\bar{z}}+1}\right)\right)+ O(z^{\frac{3}{2}}).
\end{split}
\end{equation}
Inserting these discontinuities into \eqref{defectinversionseries} we find
\begin{equation}\label{Bnonid}
\begin{split}
    &B_{S}(z,\beta)_{\text{not id}}= \frac{3 \sqrt{z} \left(2 (\beta +1)-2 (\beta -1) \beta  \left(\psi ^{(0)}\left(\frac{\beta }{2}\right)+\gamma \right)+(\beta -1) \beta  \left(2 H_{\frac{\beta -3}{2}}+\log (z)\right)\right)}{4(\beta -1) \beta ^2},\\
    &B_{V}(z,\beta)_{\text{not id}}= \frac{\sqrt{z} \left(2 (\beta +1)-2 (\beta -1) \beta  \left(\psi ^{(0)}\left(\frac{\beta }{2}\right)+\gamma \right)+(\beta -1) \beta  \left(2 H_{\frac{\beta -3}{2}}+\log (z)\right)\right)}{4(\beta -1) \beta ^2}.
\end{split}
\end{equation}
where $H_z$ are harmonic numbers. Combining both contributions \eqref{Bid} and \eqref{Bnonid} in \eqref{defectinversionseries} we obtain
\begin{equation}\label{defectcoeff}
\begin{split}
     b_{S} (\hat{\Delta},s) &\sim \frac{s-1}{(2 s+1) (\hat{\tau} -1)^2}+\frac{2 (s-1) H_s+3 H_{s+\frac{1}{2}}}{2(2s+1)(\hat{\tau}-1)}, \\
      b_{V} (\hat{\Delta},s) &\sim \frac{s}{(2 s+1) (\hat{\tau} -1)^2}+\frac{(2 s+1) \left(2 s H_s+H_{s-\frac{1}{2}}\right)+2}{2 (2 s+1)^2 (\hat{\tau}-1)}.
\end{split}
\end{equation}
The appearance of double poles signals the presence of anomalous dimensions, indeed
\begin{equation}
    b (\hat{\Delta},s) \sim \frac{\hat{b}^{(0)}+ \epsilon \hat{b}^{(1)}}{\hat{\tau}^{(0)}+\epsilon \hat{\gamma}^{(1)}   -1} = \frac{\hat{b}^{(1)}}{\hat{\tau}^{(0)} -1}-\frac{\hat{b}^{(0)} \hat{\gamma}^{(1)} }{(\hat{\tau}^{(0)} -1)^2}.
\end{equation}
Comparing with \eqref{defectcoeff}, we find the one-loop defect data
\begin{align}
         &\hat{\gamma}_{S,0,s}^{(1)}= \frac{1-s}{(2 s+1)}, &
        &\hat{b}_{S,0,s}^{2 (1)} = \frac{-2 (s-1) H_s-3 H_{s+\frac{1}{2}}}{2(2s+1)} \nonumber,\\
        &\hat{\gamma}_{V,0,s}^{(1)}= - \frac{s}{(2 s+1)}, &
        &\hat{b}_{V,0,s}^{2 (1)} = -\frac{(2 s+1) \left(2 s H_s+H_{s-\frac{1}{2}}\right)+2}{2 (2 s+1)^2}.\label{defectdata1}
\end{align}
We can perform several sanity checks on these results. First of all, we can check that $\hat{\gamma}_{S,0,1}^{(1)}=0$, i.e. the presence of the displacement operator in the defect OPE of the fundamental field. The displacement operator is a protected operator, which is related to the explicit breaking of translational symmetry and, for a line defect, it has dimension two. Furthermore, it has orthogonal spin one and it is a singlet under internal symmetries. The only candidate in this case is the operator $\partial_{\perp} \hat{\phi}^1$ which is indeed associated to the vanishing anomalous dimension $\hat{\gamma}_{S,0,1}^{(1)}$. 

Another universal protected operator is the so-called tilt operator which appears whenever the defect breaks part of the internal symmetry of the bulk theory. In this case the defect breaks $O(N)$ to $O(N-1)$ and each of the $N-1$ broken generators is associated to a tilt operator forming a vectorial representation of the preserved subgroup. This vector has dimension one and orthogonal spin zero, i.e. it is given precisely by $\hat{\phi}^{\hat i}$, the $N-1$ scalars that are not involved in the construction of the defect. Correspondingly, we find $\hat{\gamma}_{V,0,0}^{(1)}=0$. 

For other defect operators some results are already avaliable in the literature. The anomalous dimensions for the spin-zero singlet, i.e. $\hat \phi^1$ and the spin-one vector operators, i.e. $\partial_{\perp}\hat \phi^{\hat i}$, have been computed in equations (3.19) and (3.52) of \cite{Cuomo:2021kfm} and we find perfect agreement. More generally, the leading-twist contribution to the defect anomalous dimensions from the inversion of a single bulk scalar operator of arbitrary dimension $\Delta$ was computed in (3.53) of \cite{Lemos:2017vnx}. The perturbative expansion of that result perfectly matches ours. Our results for the bulk-to-defect couplings are completely new.

As we already mentioned, the one-loop result contains information about the anomalous dimensions of operators that already appeared at tree level. Nevertheless, keeping more terms in the small-$z$ expansion \eqref{defectinversionseries} and \eqref{disczseries} we can also extract the bulk-to-defect couplings for higher-twist operators. As one could have guessed from a diagrammatic argument and as we anticipated in \eqref{defectopepert} it turns out that all higher-twist coefficients are zero at this order in perturbation theory. 
\begin{align}
        &\hat{b}_{S,m,s}^{2 (1)} =
        \hat{b}_{V,m,s}^{2 (1)} =0, \quad m>0.
\end{align}
Therefore the only non zero OPE data in the defect channel are \eqref{defectdata1} and the defect identity at one loop \eqref{1ptphi}.

\subsection{Bulk channel data}\label{bulkdata}
Finally, we can extract the bulk data using the bulk inversion formula \eqref{bulkinversion}. The bulk anomalous dimensions are well-known and reproducing them is just a consistency check for the validity of our procedure. On the other hand, the product $a \lambda^{(1)}$ depends on the defect through the one-point function $a$ and therefore we can provide new predictions for all the operators not afflicted by perturbative degeneracies (in this case all the twist-two operators and the first two operator in the twist-four family). \\
In order to use the formula \eqref{bulkinversion}, we need to analyse the behaviour of $\left(\frac{(1-z)(1-\bar{z})}{\sqrt{z \bar{z}}}\right)^{\Delta_\phi} F(z, \bar{z})$ for $w\to 0$ according to \eqref{smallwl} \footnote{Notice that for the bulk inversion we need to consider an extra factor in front of the correlator $F(r,w)$. For this reason the behaviour at small $w$ is different from the one discussed in the defect inversion section.}. In our case, we can see that $\ell_{*}=2$ by expanding the results of the dispersion relation \eqref{F1S} and \eqref{F1T} around $w=0$ and comparing with \eqref{smallwl}, therefore the inversion formula will work for $\ell > 2$. \\
Since the integral in \eqref{bulkinversion} is too hard, we follow the same strategy that we used in the case of the defect inversion. We expand the integrand around $z=1$ and compute the coefficient term by term in the expansion. Namely we use \cite{Liendo:2019jpu}
\begin{equation}\label{bulkinversionseries}
    \begin{split}
     &c^{t}(\Delta,\ell) = \int_{0}^{1}  \frac{dz}{2(1-z)} (1-z)^{\frac{\ell - \Delta}{2}} \sum_{m=0} (1-z)^{m} \sum_{k=-m}^{m} B_{m,k}(\Delta,\ell) C^{t}(z,\Delta+\ell+2k), \\
     & C^{t}(z,\beta) = \kappa_{\beta} \int_{0}^{z} \frac{d \bar{z}}{(1-\bar{z})^2}  k_{\beta}(1-\bar{z}) \text{dDisc}\left(\left(\frac{(1-z)(1-\bar{z})}{\sqrt{z \bar{z}}}\right)^{\Delta_\phi} F(z,\bar{z})\right), \\
     &k_{\beta}(z)= z^{\frac{\beta}{2}} {}_2 F_{1} (\frac{\beta}{2},\frac{\beta}{2},\beta,z),
\end{split}
\end{equation}
where $B_{m,k}(\Delta,\ell)$ are coefficients that can be fixed by expanding the kernel in \eqref{bulkinversion} and comparing with \eqref{bulkinversionseries}, namely
\begin{equation}
\begin{split}
  &(1-z) (1-\bar{z})^2 (1-z)^{\frac{\Delta -l}{2}} \mu (z,\bar{z}) f^{HS}_{d+l-1,-d+\Delta +1}= \\
  &=\sum _{m} (1-z)^m \sum _{k=-m}^m \frac{ \kappa_{\Delta +2 k+l} }{\kappa_{\Delta +l}}B_{m,k}(\Delta,\ell) k_{\Delta+\ell+2k}(1-\bar{z}).
\end{split}
\end{equation}
Just like in the defect case, a given term in the series expansion around $z=1$ \eqref{bulkinversionseries} reproduces the contribution of a given twist to the coefficient. \\
We can compute the double discontinuity at order $\epsilon$ by taking the expressions for the full correlator \eqref{F1S} and \eqref{F1T}, multiplying by $\frac{(1-z)(1-\bar{z})}{\sqrt{z \bar{z}}}$ and expanding in $z=1$. Every term in the expansion of the non trivial part of the correlator $ \frac{(1-z)(1-\bar{z})}{\sqrt{z \bar{z}}} I(r,w)$ contains only integer powers of $\bar{z}$.
Therefore, we see that this part of the correlator does not contribute to the double discontinuity. 
The same applies for the contribution from the correction to the bulk identity. In other words, the only contributions to the double discontinuity of both representations at one loop are the terms proportional to the defect identity, ${a^{2}_{\phi}}^{(1)} \frac{(1-z)(1-\bar{z})}{\sqrt{z \bar{z}}}$. However, the one-loop coefficient \eqref{bulkinversionseries} will also receive a contribution from the tree-level double discontinuity because it contains factors that depend on $\Delta_{\phi}$ and $d$, which will give terms of order $\epsilon$ when combined with the order zero double discontinuity \eqref{disc0}. In particular both $C^{t}(z, \beta)$ and $B_{m,k}(\Delta,\ell)$ depend on $\epsilon$ through $\Delta_\phi$ and $d$.  All in all, we need to consider
\begin{equation}
\begin{split}
    & \textstyle \text{dDisc}\left( \left(\frac{(1-z)(1-\bar{z})}{\sqrt{z \bar{z}}}\right)^{\Delta_\phi} F_S (r,w)\right) = \left( \frac{{a^{2}_{\phi}}^{(0)}}{N}+ \epsilon \frac{{a^{2}_{\phi}}^{(1)}}{N}\right)  2 \sin ^2\left(\frac{\pi  \Delta_{\phi} }{2}\right)\left(\frac{(1-z)(1-\bar{z})}{\sqrt{z \bar{z}}}\right)^{\Delta_\phi}, \\
    & \textstyle \text{dDisc} \left( \left(\frac{(1-z)(1-\bar{z})}{\sqrt{z \bar{z}}}\right)^{\Delta_\phi} F_T (r,w)\right) =\left( {a^{2}_{\phi}}^{(0)}+ \epsilon {a^{2}_{\phi}}^{(1)}\right)  2 \sin ^2\left(\frac{\pi  \Delta_{\phi} }{2}\right)\left(\frac{(1-z)(1-\bar{z})}{\sqrt{z \bar{z}}}\right)^{\Delta_\phi}.
\end{split}
\end{equation}
We computed the first few contributions to $C^{t}(z, \beta)$ in the $z=1$ expansion in the appendix \eqref{Ct1loop}, for both representations.
Plugging those expressions in \eqref{bulkinversionseries}, using \eqref{1ptphi} and expanding at first order in $\epsilon$, we can find the one-loop coefficients $c_{S/T}^t (\Delta,\ell)$. Since the external operators are identical, the u-channel contribution $c_{S/T}^u (\Delta,\ell)$ is the same. At the end of the day, we can extract the one-loop bulk OPE data for both representations
\begin{equation}\label{bulkdata1}
    \begin{split}
         &a \lambda^{(1)}_{T,0,\ell}= - \frac{2^{-\ell-5} \Gamma \left(\frac{\ell}{2}+\frac{1}{2}\right)^3}{\pi  \Gamma \left(\frac{\ell}{2}+1\right) \Gamma \left(\ell+\frac{1}{2}\right)} \Big(-32 {a^{2}_{\phi}}^{(0)} H_{\frac{\ell}{2}-\frac{1}{2}}+35 {a^{2}_{\phi}}^{(0)} H_{\ell-\frac{1}{2}}+19 {a^{2}_{\phi}}^{(0)} \psi ^{(0)}(\ell)+ \\ &-38 {a^{2}_{\phi}}^{(0)} \psi ^{(0)}(2 \ell)-19 \gamma  {a^{2}_{\phi}}^{(0)}+6{a^{2}_{\phi}}^{(0)} \log (2)+32 {a^{2}_{\phi}}^{(1)}\Big),\\
         &a \lambda^{(1)}_{T,1,\ell}= -\frac{{a^{2}_{\phi}}^{(0)} 2^{-\ell-3} \Gamma \left(\frac{\ell+1}{2}\right) \Gamma \left(\frac{\ell+3}{2}\right)^2}{\pi \Gamma \left(\frac{\ell}{2}+2\right) \Gamma \left(\ell+\frac{3}{2}\right)}, \\
         &a \lambda^{(1)}_{S,0,\ell}= \frac{1}{N} \lambda^{(1)}_{T,0,\ell}, \\
         &a \lambda^{(1)}_{S,1,\ell}= \frac{1}{N} \lambda^{(1)}_{T,1,\ell}.
    \end{split}
\end{equation}
The anomalous dimensions are all zero for $\ell>0$, as we already knew.
We also computed $C^{t}(z, \beta)$ and \eqref{bulkinversionseries} to higher order in the $z=1$ expansion to extract the coefficients and anomalous dimensions of higher-twist operators and checked that they turn out to be zero. 
The data at low spin  have to be computed in a different way, because the inversion formula does not converge in that case. For example we can compute the missing data by expanding the full results \eqref{F1S} and \eqref{F1T} in series and comparing it to the bulk OPE expansion. Although we expected the inversion formula to fail at spin $\ell=0,2$, we find out that the only data that are missed by the inversion formula are the coefficients of the twist-two operators with $\ell=0$ \eqref{bulkchannelexpansion}. We obtain
\begin{equation}
\begin{split}
    &a\lambda^{(1)}_{T,0,0}= {a^{2}_{\phi}}^{(1)}-\frac{1}{2}-\frac{\log (2)}{2},\\
    &a\lambda^{(1)}_{S,0,0}= -\frac{-4 {a^{2}_{\phi}}^{(1)}+N+N \log (2)+2+\log (4)}{4 N}.
\end{split}
\end{equation}
As another non-trivial check of our results at twist $4$ at low spin, we can follow an alternative route to derive the coefficient of the operator $\phi^4$, namely $a\lambda^{(1)}_{S/T,1,0}$, and compare with the result from the inversion formula. For the sake of simplicity we set $N=1$ and therefore consider only the singlet representation. We can take the two-point function of the operator $\phi^2$ at tree level and extract the bulk CFT data. Just like in the case of the two-point function of $\phi$, the tree-level correlator has a contribution from the squared one-point function and from the bulk identity with dimension $\Delta_{\phi^2}=2+O(\epsilon)$, namely
\begin{equation}
    F^{(0)}(r,w)= {a^{2}_{\phi^2}}^{(0)}+{a^{2}_{\phi}}^{(0)}\frac{r w}{(r-w) (r w-1)}+\left(\frac{r w}{(r-w) (r w-1)}\right)^2.
\end{equation}
A family of double-twist operators with twist $4$ enters in the bulk spectrum at this order. In particular, we can extract the coefficient  of the spin zero double twist $a \lambda^{(0)}_{\phi^2 \phi^2 \phi^4 }= a^{(0)}_{\phi^4} \lambda^{(0)}_{\phi^2 \phi^2 \phi^4}$ from the bulk expansion. Since the squared three-point coefficients $\lambda^{2 \ (0)}_{\phi^2 \phi^2 \phi^4}$ are known from previous work on the theory without defects  \cite{Henriksson:2022rnm}, one can immediately derive an expression for the squared one-point function $a^{2 \ (0)}_{\phi^4}$ from $a \lambda^{(0)}_{\phi^2 \phi^2 \phi^4 }$. Multiplying this expression and the known squared one-loop OPE coefficients $\lambda^{2 \ (1)}_{\phi \phi \phi^4 }$, one can compute the square of the $\phi^4$ coefficient $a^{2 \ (0)}_{\phi^4} \lambda^{2 \ (1)}_{\phi \phi \phi^4 }=a \lambda^{2 (1)}_{S,1,\ell}$, where we used the fact that  $\lambda^{2 \ (0)}_{\phi \phi \phi^4 }=0$. The results we obtain match with what we expect from \eqref{bulkdata1}. Obviously this reasoning can be extended to any $N$. Considering other correlators is not only useful to check results but also to solve the mixing problem. Indeed, starting at spin $\ell=2$, the twist-four operators suffer from degeneracies, namely there are multiple operators that have the same classical scaling dimensions and therefore are indistinguishable from the bootstrap perspective. For this reason the data that we found at twist four \eqref{bulkdata1} has to be interpreted as an average over all degenerate twist-four operators for a given spin. In order to lift the degeneracy, one should in principle compare the results from different correlators.

\section*{Acknowledgments} 
We would like to thank A. Gimenez-Grau for coordinating the submission of his work with ours. We are also grateful to M. Billò, A. Bissi, G. Bliard, V. Forini, F. Galvagno, P. Garau, M. Meineri, M. Lemos, P. van Vliet for useful discussions. The research of LB is funded through the MIUR program for young researchers “Rita Levi Montalcini”. The research of DB received partial support through the STFC grant ST/S005803/1.

\appendix

\section{Kinematics and conformal blocks}
\label{app:kinematics}
In this work we consider the two-point function
\begin{equation}
    \braket{\phi_i(x) \phi_j(y)} = \frac{F_1(z,\bar{z})\delta_{ij}+F_2(z,\bar{z})\delta_{i1}\delta_{j1}}{|x_{\perp}|^{\Delta_{\phi}}|y_{\perp}|^{\Delta_{\phi}}},
\end{equation}
where $z$ and  $\bar{z}$ are the lightcone coordinates defined by
\begin{equation}
\xi_1=\frac{(x-y)^2}{4|x_\perp||y_\perp|}, \quad \xi_2=\frac{x\cdot 
 y}{|x_\perp||y_\perp|}; \quad \xi_1=\frac{(1-z)(1-\bar{z})}{4\sqrt{z\bar{z}}}, \quad \xi_2=\frac{z+\bar{z}}{2\sqrt{z\bar{z}}}.
\end{equation}
They  are related to  the radial coordinates by
\begin{equation}
z=rw, \quad \bar{z}=\frac{r}{w}.
\end{equation}
The functions $F_1(z,\bar{z})$ and $F_2(z,\bar{x})$ can be expressed as linear combinations of $F_S(z,\bar{z})$, $F_T(z,\bar{z})$ and of $\hat{F}_S(z,\bar{z})$, $\hat{F}_T(z,\bar{z})$ defined in \eqref{FSFT} and \eqref{FSFV}
\begin{equation}\label{linearcombinations}
\begin{split}
&F_1(z,\bar{z})=F_S(z,\bar{z})-\frac{1}{N}F_T(z,\bar{z})=\hat{F}_V(z,\bar{z}),\\
&F_2(z,\bar{z})=F_T(z,\bar{z})=\hat{F}_S(z,\bar{z})-\hat{F}_V(z,\bar{z}).
\end{split}
\end{equation}
The functions $F_1(z,\bar{z})$ and $F_2(z,\bar{z})$ obey crossing equations
\begin{equation}
  \begin{split}
    F_1(z,\bar{z}) = \left(\frac{\sqrt{z \bar z}}{(1-z)(1-\bar z)}\right)^{\Delta_{\phi}} \sum_{\Delta, \ell} \,\bigg( a_{\mathcal{O}_S} \,&\lambda_{\phi\phi\mathcal{O}_S}-   \frac{1}{N}a_{\mathcal{O}_T} \,\lambda_{\phi\phi\mathcal{O}_T}\bigg)\, f_{\Delta,\ell}(z,\bar{z})=\\
 & \ \ \ \ \ \ \ \ \ \ \ \ \ = \sum_{\hat{\Delta},s}\, b_{V, \hat{\Delta},s}^2 \, \hat{f}_{\hat\Delta,s}(z,\bar{z})\,, \\
      F_2(z,\bar{z}) = \left(\frac{\sqrt{z \bar z}}{(1-z)(1-\bar z)}\right)^{\Delta_{\phi}} \sum_{\Delta, \ell} \, a_{\mathcal{O}_T} \, & \lambda_{\phi\phi\mathcal{O}_T}\, f_{\Delta,\ell}(z,\bar{z})=\\
     & =\sum_{\hat{\Delta},s}\,\left( b_{S, \hat{\Delta},s}^2-b_{V, \hat{\Delta},s}^2 \right)\, \hat{f}_{\hat\Delta,s}(z,\bar{z})\,, \\
  \end{split}
\end{equation}
where the defect conformal blocks are given by
\begin{equation}
    \hat{f}_{\hat\Delta,s}(z,\bar{z}) = z^{\frac{\hat{\Delta}-s}{2}} \bar{z}^{\frac{\hat{\Delta}+s}{2}} {}_2 F_{1} \left(-s, \frac{q}{2}-1,2-\frac{q}{2}-s,\frac{z}{\bar{z}}\right) {}_2 F_{1}\left(\hat{\Delta},\frac{p}{2},\hat{\Delta}+1-\frac{p}{2},z \bar{z}\right),
\end{equation}
with $p=1$ and $q=d-1$ for our case.
They factorize in radial coordinates
\begin{equation}
        \hat{f}_{\hat\Delta,s}(r,w) =\hat f_{\hat \Delta}(r) \hat g_{s}(w) ,
\end{equation}
with 
\begin{align}
\hat f_{\hat \Delta}(r)&= r^{\hat{\Delta}} \, {}_2 F_1 \left(\hat{\Delta},\frac{p}{2},\hat{\Delta}+1-\frac{p}{2},r^2\right)\, , & \hat g_{s}(w)=w^{-s} {}_2 F_1 \left(-s,\frac{q}{2}-1,2-\frac{q}{2}-s,w^2\right).
\end{align}
The bulk blocks are not known in a closed form, but can be expressed as a sum of Harish-Chandra functions \cite{Isachenkov:2018pef}
\begin{equation}
    f_{\Delta,\ell}(z,\bar{z})= f^{HS}_{\Delta,\ell}(z,\bar{z})+\frac{\Gamma(\ell+d-2)\Gamma(-\ell-\frac{d-2}{2})}{\Gamma(\ell+\frac{d-2}{2})\Gamma(-\ell)} \frac{\Gamma(\frac{\ell}{2}+\frac{d-p}{2}-\frac{1}{2})\Gamma(-\frac{\ell}{2}+\frac{1}{2})}{\Gamma(\frac{\ell}{2}+\frac{d}{2}-\frac{1}{2})\Gamma(-\frac{\ell}{2}-\frac{p}{2}+\frac{1}{2})}f^{HS}_{\Delta,2-d-\ell}(z,\bar{z}),
\end{equation}
where $f_{\Delta,\ell}^{HS}(z,\bar{z})$ can be expressed as a double infinite sum
\begin{equation}
\begin{split}\label{bulkblock}
        f_{\Delta,\ell}^{HS}(z,\bar{z}) &=\sum_m^{\infty}\sum_n^{\infty} [(1-z)(1-\bar{z})]^{\frac{\Delta-\ell}{2}+m+n}
        h_n (\Delta,\ell)h_m (1-\ell,1-\Delta) \frac{4^{m-n}}{n! m!} \frac{(\frac{\Delta+\ell}{2})_{n-m}}{\left(\frac{\Delta+\ell}{2}-\frac{1}{2} \right)_{n-m}}  \\ &\textstyle \times {}_4 F_3 (-n,-m,\frac{1}{2},\frac{\Delta-\ell}{2}-\frac{d}{2}+1;-\frac{\Delta+\ell}{2}+1-n,\frac{\Delta+\ell}{2}-m,\frac{\Delta-\ell}{2}-\frac{d}{2}+\frac{3}{2};1) \\
        & (1-z \bar{z})^{\ell-2m}  \textstyle {}_2 F_1 (\frac{\Delta+\ell}{2}-m+n,\frac{\Delta+\ell}{2}-m+n,\Delta+\ell-2(m-n),1-z \bar{z}),
\end{split}
\end{equation}
where
\begin{equation}
    h_n (\Delta,\ell)=\frac{\left(\frac{\Delta}{2}-\frac{1}{2},\frac{\Delta}{2}-\frac{p}{2},\frac{\Delta+\ell}{2} \right)_n }{\left( \Delta-\frac{d}{2}+1,\frac{\Delta+\ell}{2}+\frac{1}{2}\right)_n}.
\end{equation}

\section{Explicit computation of the one-loop correlator}
\label{app:explicit}
As we discussed in the main text, the discontinuity of the correlator contains a non trivial term that gives rise to a very complicated integral in the dispersion formula. Here we will show the computation of that integral and in particular we will give a derivation of $I(r,w)$ \eqref{correlatormeijer}, namely the non trivial contribution to the full correlators \eqref{F1S} \eqref{F1T} .
We start by writing the non trivial part of the discontinuity \eqref{discellipticS} as a sum of hypergeometric functions, using the definition of the bulk block \eqref{bulkblock}, then we replace the hypergeometric function with its definition as a sum and finally we exchange the order of the sums. We find 
\begin{equation}
\begin{split}
     &\frac{4 \pi  \sqrt{r w} \left(F\left(\sin ^{-1}\left(\sqrt{r} \sqrt{\frac{r-w}{1-r w}}\right)|\frac{(r w-1)^2}{(r-w)^2}\right)-F\left(\sin ^{-1}\left(\frac{\sqrt{\frac{r-w}{1-r w}}}{\sqrt{r}}\right)|\frac{(r w-1)^2}{(r-w)^2}\right)\right)}{r-w} =\\ &=  \sum_{n=0}^{\infty}\frac{i \pi  2^{1-2 n} r}{1+2n} \left(\frac{(r-w) (r w-1)}{w}\right)^n \, {}_2 F_1\left(n+1,n+1;2 n+2;1-r^2\right)=\\
     &=\sum_{n=0}^{\infty} \sum_{m=0}^{\infty} \frac{i \pi  2^{1-2 n} r}{1+2n} \left(\frac{(r-w) (r w-1)}{w}\right)^n \frac{(1-r^2)^m \left((n+1)_m (n+1)_m\right)}{m! (2 n+2)_m}=\\
    &= \sum_{m=0}^{\infty} \frac{2 i \pi  r \left(1-r^2\right)^m}{m+1} {}_3F_2\left(\frac{1}{2},m+1,m+1;\frac{m}{2}+1,\frac{m}{2}+\frac{3}{2};\frac{(r-w) (r w-1)}{4 w}\right).
\end{split}
\end{equation}
It turns out that this expansion of the discontinuity can be easily integrated term by term  in the dispersion relation and we can find the function $I(r,w)$. Indeed, plugging the previous expression into the dispersion relation \eqref{dispersion} and changing variable from $w'$ to $x=\Big(\frac{(1-r w')(w'-r)}{4 w'}\Big)$, we find 
\begin{equation}\label{correlatormeijer1}
    \begin{split}
        \textstyle I(r,w)&= \int_{-\infty}^0 dx \  \sum_{m=0}^{\infty}  \frac{4r w (1-r^2)^m \, _3F_2\left(\frac{1}{2},m+1,m+1;\frac{m}{2}+1,\frac{m}{2}+\frac{3}{2};x\right)}{(m+1) \left(r^2 (-w)+r w^2+r+4 w x-w\right)}= \\
        &= \sum_{m=0}^{\infty}  \frac{2^{1-m} r w (1-r^2)^m G_{4,4}^{4,2}\left(\frac{4 w}{w r^2-\left(w^2+1\right) r+w}|
\begin{array}{c}
 0,0,\frac{m}{2},\frac{m+1}{2} \\
 -\frac{1}{2},0,m,m \\
\end{array}
\right)}{(m!)\left(r^2 (-w)+r w^2+r-w\right)},
    \end{split}
\end{equation}
where $G$ is the Meijer G-function and we assumed
\begin{equation}\label{conditionmeijer}
    \Re\left(\frac{r \left(-r w+w^2+1\right)}{w}\right)\leq 1\lor \frac{r \left(-r w+w^2+1\right)}{w}\notin \mathbb{R}.
\end{equation}
To rewrite the previous expression in a closed form we can use the integral representation of the Meijer G-function
\begin{equation}\label{Meijer}
    G_{p,q}^{m,n}\left(z\left|
\begin{array}{c}
 a_1,\ldots ,a_n,a_{n+1},\ldots ,a_p \\
 b_1,\ldots ,b_m,b_{m+1},\ldots ,b_q \\
\end{array}
\right.\right)=\frac{1}{2 \pi  i} \int_{\mathcal{L}} ds \ \frac{\prod _{k=1}^n \Gamma \left(-s-a_k+1\right) \prod _{k=1}^m \Gamma \left(s+b_k\right) z^{-s}}{\prod _{k=n+1}^p \Gamma \left(s+a_k\right) \prod _{k=m+1}^q \Gamma \left(-s-b_k+1\right)}.
\end{equation}
Plugging the integral representation representation in \eqref{correlatormeijer1}, and exchanging sum and integral, we find
\begin{equation}
    \begin{split}
        &\textstyle I(r,w) =  \sum\limits_{m=0}^{\infty} \int ds \ \frac{2^{1-m} r w \left(1-r^2\right)^m}{\Gamma (m+1) (w-r) (r w-1)} \frac{2^{m+1} \Gamma (1-s)^2 \Gamma (2 s-1) \Gamma (m+s)^2}{\Gamma (m+2 s)} \left(\frac{4 w}{(r-w) (r w-1)}\right)^{-s}=\\
        &= - \int_{\mathcal{L}} ds \ \frac{\pi ^2 r 4^{1-s} \csc ^2(\pi  s)}{2 s-1} {}_2F_1\left(s,s;2 s;1-r^2\right) \left(\frac{w}{(r-w) (r w-1)}\right)^{1-s}= \\
        &= \textstyle  \sum\limits_{n=0}^{\infty}  \frac{\partial}{\partial t}\left( \left(\frac{(r-w) (r w-1)}{w}\right)^{n+t} \frac{4^{-n-t} r }{2n+2t+1} \, _2F_1\left(n+t+1,n+t+1;2 (n+t+1);1-r^2\right) \right)= \\
        &= \textstyle  \sum\limits_{n=0}^{\infty}  \frac{\partial }{\partial t} \left( \left(\frac{(r-w) (r w-1)}{(1+r)^2 w}\right)^{n+t} \frac{4 r}{(1+r)^2 (2n+2 t+1)} {}_2F_1\left(\frac{1}{2},n+t+1;n+t+\frac{3}{2};\frac{(1-r)^2}{(r+1)^2}\right) \right)=\\
         &= \left. \textstyle \frac{\partial }{\partial t} \left( \left(\frac{(r-w) (r w-1)}{(1+r)^2 w}\right)^t \frac{4 r}{(1+r)^2 (2 t+1)} F^{112}_{ 101}\left(
\begin{array}{c}
 1+t:\frac{1}{2};\frac{1}{2}+t,1; \\
 \frac{3}{2}+t:-;1+t; \\
\end{array}
\ \left(\frac{1-r}{1+r}\right)^2,\frac{(r-w) (r w-1)}{(1+r)^2 w}\right) \right)\right|_{t=0}.
    \end{split}
\end{equation}
The final result is expressed in terms of a derivative of a Kampé de Fériet function \cite{KdFconvergence}, which is defined by 
\begin{equation}\label{kdfdef}
    F^{ABB'}_{CDD'}\left(
    \begin{array}{c}
     (a):(b);(b'); \\
     (c):(d);(d'); \\
    \end{array}
    \ x,y\right) = \sum_{m=0}^{\infty} \sum_{n=0}^{\infty} \frac{\prod_{j=1}^{A} (a_j)_{m+n} \prod_{j=1}^{B} (b_j)_{m} \prod_{j=1}^{B'} (b'_j)_{n}}{\prod_{j=1}^{C} (c_j)_{m+n} \prod_{j=1}^{D} (d_j)_{m} \prod_{j=1}^{D'} (d'_j)_{n}} \frac{x^m}{m!}\frac{y^n}{n!}.
   \end{equation}
   This sum converges absolutely if $A+B=C+D+1$, $A+B'=C+D'+1$, $|x|<1$ and $|y|<1$, which in our case, together with \eqref{conditionmeijer}, imply
    \begin{equation}
       0<r<1\land \left((0<w<1\land 0<r\leq w)\lor (w=1\land 0<r<1)\lor \left(w>1\land 0<r\leq \frac{1}{w}\right)\right).
    \end{equation}
    We can extend the result to all other values by exploiting the symmetry of the correlator with respect to $r \leftrightarrow \frac{1}{r}$ and $w \leftrightarrow \frac{1}{w}$ \cite{Lemos:2017vnx}.
\subsection{Generating function of the bulk inversion}
The first few orders in the $z=1$ expansion of the one-loop generating function $C^{t}_T (z, \beta)$ \eqref{bulkinversionseries} read
\begin{equation}\label{Ct1loop}
\begin{split}
   &C^{t}_T (z, \beta) =  {a^{2}_{\phi}}^{(0)} \Bigg((1-z)^{\Delta_{\phi} } \Bigg(\frac{\sqrt{\pi } \Gamma \Big(\frac{\beta +1}{2}\Big) \Gamma \Big(1-\frac{\Delta_{\phi} }{2}\Big)^2 \Gamma \Big(\frac{\beta }{2}+\Delta_{\phi} -1\Big)}{\Gamma \Big(\frac{\beta +2}{4}\Big)^2 \Gamma \Big(\frac{1}{4} (\beta -2 \Delta_{\phi} +4)\Big) \Gamma \Big(\frac{1}{4} (\beta +2 \Delta_{\phi} )\Big)} +\\ &+\Bigg(\frac{(z-1)^2 \Big(-\beta ^3-4 \beta ^2 \Delta_{\phi} -6 \beta ^2-4 \beta  \Delta_{\phi} ^2-20 \beta  \Delta_{\phi} -8 \beta -4 \Delta_{\phi} ^2-16 \Delta_{\phi} \Big)}{192 (\beta +1)}+ \\ &+\frac{(z-1) \Big(\beta ^3+4 \beta ^2 \Delta_{\phi} +4 \beta ^2+4 \beta  \Delta_{\phi} ^2+12 \beta  \Delta_{\phi} +4 \beta +4 \Delta_{\phi} ^2+8 \Delta_{\phi} \Big)}{16 (\beta +1) (\beta +2 \Delta_{\phi} +2)}+\\ &+\frac{2}{(z-1) (\beta +2 \Delta_{\phi} -2)} -\frac{1}{2}\Big) (1-z)^{\frac{\beta }{2}+\Delta_{\phi} }\Bigg) \times \\ & \times \Bigg(\frac{\Gamma \Big(\frac{\beta }{2}\Big)^4 \sin ^2\Big(\frac{\pi  \Delta_{\phi} }{2}\Big)}{\pi ^2 \Gamma (\beta -1) \Gamma (\beta )}+\frac{\Big(\Delta_{\phi} ^2+2 \Delta_{\phi} \Big) (z-1)^2 \Gamma \Big(\frac{\beta }{2}\Big)^4 \sin ^2\Big(\frac{\pi  \Delta_{\phi} }{2}\Big)}{8 \pi ^2 \Gamma (\beta -1) \Gamma (\beta )}+\\ & -\frac{\Big(\Delta_{\phi} ^3+6 \Delta_{\phi} ^2+8 \Delta_{\phi} \Big) (z-1)^3 \Gamma \Big(\frac{\beta }{2}\Big)^4 \sin ^2\Big(\frac{\pi  \Delta_{\phi} }{2}\Big)}{48 \pi ^2 \Gamma (\beta -1) \Gamma (\beta )}-\frac{\Delta_{\phi}  (z-1) \Gamma \Big(\frac{\beta }{2}\Big)^4 \sin ^2\Big(\frac{\pi  \Delta_{\phi} }{2}\Big)}{2 \pi ^2 \Gamma (\beta -1) \Gamma (\beta )}\Bigg) \Bigg) \Bigg)+ \\
    &-{a^{2}_{\phi}}^{(1)} \frac{\Gamma \left(\frac{\beta }{4}\right)^2 (1-z)^{\Delta \phi } \Gamma \left(\frac{\beta }{4}+\frac{\Delta \phi }{2}-\frac{1}{2}\right)}{\Gamma \left(\frac{\beta }{2}-\frac{1}{2}\right) \left(2^{\frac{\beta }{2}-\Delta \phi +1} \Gamma \left(\frac{\Delta \phi }{2}\right)^2\right) \Gamma \left(\frac{\beta }{4}-\frac{\Delta \phi }{2}+1\right)}+O((z-1)^4),
\end{split}    
\end{equation}
where $\Delta_\phi$ needs to be expanded at one loop.
The singlet contribution is simply $C^{t}_S (z, \beta) = \frac{1}{N} C^{t}_T (z, \beta)$.

\section{Feynman diagrams}
The bulk two-point function of two fundamental fields in the presence of the defect $\langle \phi_i(x) \phi_j(y) \rangle_D$ can be calculated perturbatively by evaluating Feynman diagrams at the fixed point. In particular, to the first order in $\epsilon$ (hence at one loop) there are two disconnected contributions, which are given by the one-loop correction to the bulk propagator and the one-loop correction to the product of one-point functions
\begin{equation}
	\begin{tikzpicture}[scale=0.8]
	 \draw[double,thick,blue] (-3,0)--(3,0);
	\draw[thick] (-1.5,3)--(-0.5,3);
	\draw[thick] (0.5,3)--(1.5,3);
	\draw[fill=gray] (0,3) circle (0.5);
	 \node[above] at (-1.5,3) {$\phi$};
     \node[above] at (1.5,3) {$\phi$};
	\end{tikzpicture}\hspace{0.5 cm}
	\begin{tikzpicture}[scale=0.8]
	 \draw[double,thick,blue] (-3,0)--(3,0);
	\draw[thick] (-1.5,3)--(-1.5,0);
	\draw[thick] (1.5,0)--(1.5,3);
	\draw[fill=gray] (-1.5,1.5) circle (0.5);
	 \node[above] at (-1.5,3) {$\phi$};
     \node[above] at (1.5,3) {$\phi$};
	\end{tikzpicture}
\end{equation}
where the blue double line at the bottom represents the defect $D$, whereas black lines are scalar propagators with a generic one-loop correction represented as a grey circle. These two contributions naturally provide the corrections to the defect and bulk identity in \eqref{F1S} and \eqref{F1T}. The only connected diagram is the cross diagram
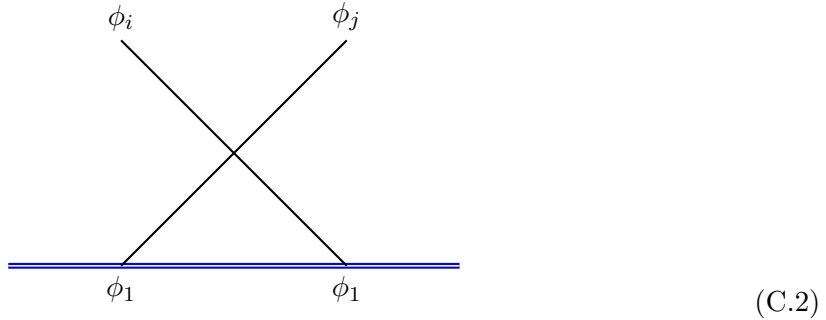
\begin{equation}\label{diagram1}
	\begin{tikzpicture}
	 \draw[double,thick,blue] (-3,0)--(3,0);
	 \draw[thick] (-1.5,3)--(1.5,0);
	 \draw[thick] (-1.5,0)--(1.5,3);
	 \node[above] at (-1.5,3) {$\phi_i$};
     \node[above] at (1.5,3) {$\phi_j$};
     \node[below] at (1.5,0) {$\phi_1$};
    \node[below] at (-1.5,0) {$\phi_1$};
	\end{tikzpicture}
\end{equation}
The contribution of the diagram \eqref{diagram1} is given by
\begin{equation}\label{diagram1integral}
-\mathcal{N}_{\phi}^{\,2} \left(\delta_{ij}+2\delta_{i1}\delta_{j1}\right)\frac{\lambda h^2}{2!}\int d \tau_1 \int d \tau_2 \int d^{4-\epsilon}q \ G(x-q)G(y-q)G(q-x(\tau_1))G(q-x(\tau_2)),
\end{equation}
where $|\dot{x}(\tau_{1,2})|=1$, $\mathcal{N}_{\phi}$ is chosen to normalize the bulk  two-point function in the absence of the defect, and $G(x)$ is the free propagator
\begin{equation}\label{fprop}
G(x)=\frac{1}{(d-2)\Omega_{d-1}|x|^{d-2}}=\frac{1}{4\pi^2|x|^2}+O(\epsilon), \quad \mathcal{N}_\phi=2\pi+O(\epsilon).
\end{equation}
Since at the fixed point $\lambda_* \propto \epsilon$, we can set $\epsilon=0$ in all the other terms. Evaluating at the fixed point, Inserting \eqref{fprop} into \eqref{diagram1integral} and performing the $\tau_2$ and $\tau_2$ integrals one gets
\begin{equation}\label{generalintegral}
-\left(\delta_{ij}+2\delta_{i1}\delta_{j1}\right)\frac{3}{8\pi^2}\int d^4 q \, \frac{1}{|q_\perp|^2\,|x-q|^2\,|y-q|^2}=\left(\delta_{ij}+2\delta_{i1}\delta_{j1}\right)\frac{I(x,y)}{8|x_\perp||y_\perp|},
\end{equation}
where $I(x,y)$ is exactly the conformally invariant function defined in \eqref{correlatormeijer}, but expressed in the coordinates $(x,y)$. Thanks to conformal invariance, it is enough to evaluate $I(x,y)$ for $x_\parallel=y_\parallel=0$. However, the integral in \eqref{generalintegral} is very hard to solve analytically and a general closed form has not been found. Nevertheless, it is still possible to reduce it to a more useful one-dimensional integral representation which can be used to obtain series expansions and other analytic considerations. Moreover, it is also possible to obtain an exact result for particular regimes, such as $x_\perp^2+y_\perp^2=1$ or $x_\perp \parallel y_\perp$ (in lightcone coordinates, $z\bar{z}=1$ and $z=\bar{z}$ respectively).

We begin with the derivation of the integral representation. The starting point is
\begin{equation}
I(x,y)=-\frac{3}{\pi^2} \int d^3 q_\perp \int d q_\parallel  \, \frac{|x_\perp| |y_\perp|}{|q_\perp|^2\,(q_\parallel^2+|x_\perp-q_\perp|^2)\,(q_\parallel^2+|y_\perp-q_\perp|^2)}.
\end{equation}
One can introduce a Feynman parameter $\alpha$ for the last two factor in the denominator
\begin{equation}
I(x,y)=-\frac{3}{\pi^2} \int_0^1 d \alpha  \int d^3 q_\perp \int d q_\parallel  \, \frac{|x_\perp| |y_\perp|}{|q_\perp|^2\,\left(q_\parallel^2+\alpha |x_\perp-q_\perp|^2 +(1-\alpha)|y_\perp-q_\perp|^2\right)^2}.
\end{equation}
We can now perform the $q_\parallel$ integral and rearrange to obtain
\begin{equation}
I(x,y)=-\frac{3}{2\pi}\int_0^1 d \alpha  \int d^3 q_\perp \, \frac{|x_\perp| |y_\perp|}{|q_\perp|^2\,\left(|q_\perp -\tilde{q}_\perp|^2+L^2\right)^{\frac{3}{2}}},
\end{equation}
where
\begin{equation}
\tilde{q}_\perp = \alpha x_\perp +(1-\alpha)y_\perp, \quad L^2=\alpha(1-\alpha)|x_\perp -y_\perp |^2.
\end{equation}
Introducing another Feynman parameter $\xi$ one gets
\begin{equation}
I(x,y)=-\frac{9}{4\pi}\int_0^1 d \alpha \int_0^1 d \xi \int d^3 q_\perp \, \frac{\xi^{\frac{1}{2}}\cdot |x_\perp| |y_\perp|}{\left(|q_\perp -\xi \tilde{q}_\perp|^2+\xi (1-\xi )|\tilde{q}_\perp|^2+\xi L^2\right)^{\frac{5}{2}}}.
\end{equation}
After the shift $q_\perp \rightarrow q_\perp+\xi \tilde{q}_\perp$ the integral over $q_\perp$ becomes easy and it gives
\begin{equation}
I(x,y)=-3 |x_\perp| |y_\perp|\int_0^1 d \alpha \int_0^1 d \xi \, \frac{1}{\xi^{\frac{1}{2}}\left( L^2+(1-\xi) |\tilde{q}_\perp|^2\right)}.
\end{equation}
It is now convenient to pass to lightcone coordinates. Changing variable $\xi=\eta^2$ and rearranging one gets
\begin{equation}\label{twodimintegral}
I(z,\bar{z})=-6\sqrt{z\bar{z}} \int_0^1 d \alpha \int_0^1 d \eta \, \frac{1}{\left( 1+\alpha(z\bar{z}-1)-\eta^2(1+(z-1)\alpha)(1+(\bar{z}-1)\alpha)\right)}.
\end{equation}
At this point one can obtain two different representations by integrating either in $\alpha$ or in $\eta$. Integrating over $\alpha$ one gets
\begin{equation}\label{intrep1}
I(z,\bar{z})=-6\sqrt{z\bar{z}}\int_0^1 d \eta \  \frac{\log \left[ P(z,\bar{z},\eta)+\sqrt{Q(z,\bar{z},\eta})\right]-\log \left[ P(z,\bar{z},\eta)-\sqrt{Q(z,\bar{z},\eta})\right]}{\sqrt{Q(z,\bar{z},\eta})},
\end{equation}
where $P(z,\bar{z},\eta)$ and $Q(z,\bar{z},\eta)$ are the following polynomials
\begin{equation}
\begin{split}
P(z,\bar{z},\eta)&=1+z\bar{z}-\eta^2 (z+\bar{z}), \\
Q(z,\bar{z},\eta)&=(z\bar{z}-1)^2-2\eta^2(z+\bar{z}+z\bar{z}(z+\bar{z}-4))+\eta^4 (z-\bar{z})^2.
\end{split}
\end{equation}
Alternatively, it is possible to perform the integral over $\eta$ in \eqref{twodimintegral} to get
\begin{equation}\label{intrep2}
I(z,\bar{z})=-6\sqrt{z\bar{z}} \int_0^1 d \alpha \  \frac{\tanh^{-1} \left[\frac{(1+(z-1)\alpha)(1+(\bar{z}-1)\alpha}{1+\alpha(z\bar{z}-1)}\right]}{\sqrt{(1+(z-1)\alpha)(1+(\bar{z}-1)\alpha)(1+\alpha(z\bar{z}-1))}}.
\end{equation}
Both expressions \eqref{intrep1} and \eqref{intrep2} can be used to obtain series expansions of $I(z,\bar{z})$. Moreover, from \eqref{intrep2} it is also possible to extract in a closed form the term in $I(z,\bar{z})$ proportional to $\log (1-\bar{z})$. Indeed in the limit $\bar{z}\rightarrow 1$ the argument of the hyperbolic arcotangent goes to $1$ giving a logarithmic divergence, whereas the denominator remains finite. The logarithmic term is therefore given by
\begin{equation}
I(z,\bar{z})_{\log (\bar{z}-1)}=3\sqrt{z\bar{z}} \,\log (\bar{z}-1)  \int_0^1 d \alpha \ \frac{1}{\sqrt{(1+(z-1)\alpha)(1+(\bar{z}-1)\alpha)(1+\alpha(z\bar{z}-1))}}.
\end{equation}
The result of this integral can be expressed in terms of incomplete elliptic integrals of the first kind $F(\varphi,k)$
\begin{equation}
\begin{split}
I&(z,\bar{z})_{\log (\bar{z}-1)}=\\
&=6 \sqrt{z\bar{z}} \, \log (\bar{z}-1) \left( \frac{F\left(\sin^{-1} \left(\sqrt{\frac{\bar{z}-z}{\bar{z}-1}}\right), \frac{z(\bar{z}-1)^2}{(\bar{z}-z)(z\bar{z}-1)} \right)-F\left(\sin^{-1} \left( \sqrt{\frac{\bar{z}-z}{z\bar{z}-z}}\right), \frac{z(\bar{z}-1)^2}{(\bar{z}-z)(z\bar{z}-1)} \right)}{\sqrt{(\bar{z}-z)(z\bar{z}-1)}} \right).
\end{split}
\end{equation}
\subsection{Evaluation of the integral for $z=\bar{z}$}
We now proceed to solve exactly the integral for some specific regimes. The first computation is for the case $x_\perp \parallel y_\perp$ with $x_\perp \cdot y_\perp >0$.
Again we start from the integral
\begin{equation}\label{mainintegral}
I(x,y)=-\frac{3}{\pi^2} \int d^3 q_\perp \int d q_\parallel  \, \frac{|x_\perp| |y_\perp|}{|q_\perp|^2\,(q_\parallel^2+|x_\perp-q_\perp|^2)\,(q_\parallel^2+|y_\perp-q_\perp|^2)}.
\end{equation}
Thanks to the conformal invariance of this integral, this is equivalent to compute it for $x_\perp=y_\perp$ with $|x_\perp|=1$ and $d=x_\parallel-y_\parallel$ arbitrary (without loss of generality, we assume $d>0$). Indeed if $y_\perp = \alpha x_\perp$ with $x_\parallel=y_\parallel=0$, then one can get the same cross ratios by taking $x_\perp  = y_\perp $ with $|x_\perp|=1$ and $d=\frac{1-\alpha}{\sqrt{\alpha}}$. Hence we can instead solve the integral
\begin{equation}
I(x,y)=-\frac{3}{\pi^2} \int d^3 q_\perp \int d q_\parallel  \, \frac{1}{|q_\perp|^2\,(q_\parallel^2+|x_\perp-q_\perp|^2)\,((q_\parallel^2+d)+|x_\perp-q_\perp|^2)}.
\end{equation}
Integration over $q_\parallel$ is straightforward and gives
\begin{equation}
I(x,y)=-\frac{3}{2\pi} \int d^3 q_\perp \  \frac{1}{|q_\perp|^2}\cdot \frac{1}{|q_\perp+x_\perp|}\cdot \frac{1}{|q_\perp+x_\perp|^2+\left(\frac{d}{2}	\right)^2} .
\end{equation}
We can still exploit rotational invariance to set $x_\perp=(1,0,0)^T$. For convenience, we also set $a=\frac{d}{2}$. Passing to spherical coordinates $(\rho,\theta, \phi)$ and performing the longitudinal integral over $\phi$ we are left with
\begin{equation}
I(a)=-3 \, \int_{0}^{\infty} d\rho \int_{0}^{\pi} d \theta  \frac{\sin \theta }{\sqrt{\rho^2+1+2\rho \cos \theta} \ (\rho^2+1+2\rho \cos \theta+a^2)} .
\end{equation}
We change integration variable $\theta \rightarrow t=\cos \theta$ to get
\begin{equation}
I(a)=-3 \,   \int_{0}^{\infty} d\rho \, \frac{1}{(2\rho)^{3/2}} \int_{-1}^{+1} dt \frac{1 }{\sqrt{A+t} \ (B+t)},
\end{equation}
with
\begin{equation}
A=\frac{\rho^2+1}{2\rho}, \ \ B=\frac{\rho^2+1+a^2}{2\rho}.
\end{equation}
The $t$ integral can now be explicitly performed, yielding 
\begin{equation}
I(a)=-3 \, \int_0^\infty d\rho \ \frac{\arctan{\frac{|\rho+1|}{a}}-\arctan{\frac{|\rho-1|}{a}}}{\rho\cdot a}.
\end{equation}
Using the trigonometric formula for $ \arctan u - \arctan v$ and splitting the domain of integration into $(0,1)$ and $(1,\infty)$ we get
\begin{equation}
I(a)=-\frac{3}{a}\,\left[ \int_0^1 \frac{d \rho}{\rho} \ \arctan\left( \frac{2\rho a}{a^2+1-\rho^2}\right)  +\int_1^\infty \frac{d\rho}{\rho} \ \arctan \left( \frac{2a}{a^2+\rho^2-1}\right) \right].
\end{equation}
This integration can be performed as well and it gives
\begin{equation}
\begin{split}
I(a)=&-\frac{3i}{8a}\left[ \log \left( -\frac{1}{(i+a)^2} \right)\left( \log \left( -\frac{1}{(i+a)^2} \right)+2\log \left(a(2i+a) \right) \right)+ \right. \\
&\left.-\log \left( -\frac{1}{(-i+a)^2} \right)\left( \log \left( -\frac{1}{(-i+a)^2} \right)+2\log \left(a(-2i+a) \right) \right)+ \right. \\
&+ 4\,\mathrm{Li}_2\left(-\frac{i}{-i+a}\right) -4\,\mathrm{Li}_2\left(\frac{i}{-i+a}\right)-2\,\mathrm{Li}_2\left(1+\frac{1}{(-i+a)^2}\right)+\\
& \left. -4\,\mathrm{Li}_2\left(\frac{i}{i+a}\right) +4\,\mathrm{Li}_2\left(-\frac{i}{i+a}\right) + 2\,\mathrm{Li}_2\left(1+\frac{1}{(i+a)^2}\right)
\right].
\end{split}
\end{equation}
Even though in this form the result involves complex numbers, it is of course real. We can simplify the first two lines using principal logarithm identities and functional identities for the dilogarithm. After some algebraic manipulations, massive simplifications occur, and the result is
\begin{equation}
I(a)=\frac{6}{a}\cdot\left(\frac{1}{2}\log a^2 \cdot \arctan a -\mathrm{Ti}_2\,(a)\right),
\end{equation}
where
\begin{equation}
\mathrm{Ti}_2(\,a)=\frac{\mathrm{Li}_2\,(ia)-\mathrm{Li}_2\,(-ia)}{2i}=\int_0^a dt \ \frac{\arctan t}{t}.
\end{equation}
Finally, we can express the integral \eqref{mainintegral} in terms of lightcone coordinates (recall that in this regime $z=\bar{z}$) just by making the identification $2a= \frac{1-z}{\sqrt{z}}$
\begin{equation}
I(z=\bar{z}>0)=12 \, \frac{\sqrt{z}}{1-z}\, \left(\frac{1}{2}\log \left(\frac{(1-z)^2}{4z}\right)\cdot\arctan\left(\frac{1-z}{2\sqrt{z}}\right)-\mathrm{Ti}_2\,\left( \frac{1-z}{2\sqrt{z}}\right) \right).
\end{equation}
The very same method can be used to obtain a closed form expression for $x_\perp \parallel y_\perp$ with $x_\perp \cdot y_\perp <0$, or equivalently, $z=\bar{z}<0$. At the end of the calculation one finds
\begin{equation}
\begin{split}
I(z=\bar{z}<0)&=\frac{3}{2}\,\frac{\sqrt{-z}}{1-z } \left(2 \pi ^2+8\text{Li}_2\left(\frac{2 \sqrt{-z}-\sqrt{z^2-6 z+1}}{1-z}\right)-8 \text{Li}_2\left(-\frac{2 \sqrt{-z}-\sqrt{z^2-6 z+1}}{1-z}\right)-\right. \\
&\left.-2 \text{Li}_2\left(\frac{1-z-\sqrt{z^2-6 z+1}}{\left(1-\sqrt{-z}\right)^2}\right)+2 \text{Li}_2\left(\frac{1-z-\sqrt{z^2-6 z+1}}{\left(1+\sqrt{-z}\right)^2}\right)+\right.\\
&\left.+2 \text{Li}_2\left(\frac{\left(1-\sqrt{-z}\right)^2}{1-z+\sqrt{z^2-6 z+1}}\right)-2 \text{Li}_2\left(\frac{\left(1+\sqrt{-z}\right)^2}{1-z+\sqrt{z^2-6 z+1}}\right)+\right.\\
&\left.+4 \log \left(\frac{1-z+\sqrt{z^2-6 z+1}}{2 \sqrt{-z}}\right) \left(2 \log \left(\frac{1-z}{2 \sqrt{-z}}\right)-2 \log \left(1+\frac{\sqrt{z^2-6 z+1}}{2 \sqrt{-z}}\right)-\right.\right. \\
&\left.\left.-\frac{1}{2} \log \left(\left(\frac{1-\sqrt{-z}}{1+\sqrt{-z}}\right)^2\right)\right)-\log \left(\frac{(z+1)^2}{-4 z}\right) \log \left(\left(\frac{1-\sqrt{-z}}{1+\sqrt{-z}}\right)^2\right)\right).
\end{split}
\end{equation}
\subsection{Evaluation of the integral for $z\bar{z}=1$}
At last, we also derive a closed form for the integral in the case $x_\perp^2+y_\perp^2=1$ ($z\bar{z}=1$). We start again from
\begin{equation}\label{integralll}
I(x,y)=-\frac{3}{\pi^2} \int d^3 q_\perp \int d q_\parallel  \, \frac{|x_\perp| |y_\perp|}{|q_\perp|^2\,(q_\parallel^2+|x_\perp-q_\perp|^2)\,(q_\parallel^2+|y_\perp-q_\perp|^2)}.
\end{equation}
Integration over $q_\parallel$ is straightforward and yields 
\begin{equation}
I(x,y)=-\frac{3}{\pi} |x_\perp| |y_\perp| \int d^3 q_\perp \,  \frac{1}{|q_\perp|^2}\cdot  \frac{1}{|x_\perp-q_\perp|\cdot|y_\perp-q_\perp|\cdot \left( |x_\perp-q_\perp| + |y_\perp-q_\perp|	\right)}.
\end{equation}
Now we use the partial fraction decomposition
\begin{equation}
\frac{1}{A \cdot B \cdot(A+B)}=\frac{1}{B \cdot (A^2-B^2)}-\frac{1}{A \cdot(A^2-B^2)},
\end{equation}
to get two integrals 
\begin{equation}\label{intform2}
\begin{split}
I(x,y)=-\frac{3}{\pi} |x_\perp| |y_\perp| \bigg( \int  d^3 q_\perp  & \, \frac{1}{|q_\perp|^2}\cdot  \frac{1}{|y_\perp-q_\perp| \left( |x_\perp-q_\perp|^2  -|y_\perp-q_\perp|^2	\right)}-\\
&- \int d^3 q_\perp  \, \frac{1}{|q_\perp|^2}\cdot  \frac{1}{|x_\perp-q_\perp| \left( |x_\perp-q_\perp|^2  -|y_\perp-q_\perp|^2	\right)} \bigg) ,
\end{split}
\end{equation}
The second term can be reduced to the one of the first in the following manner: let $M \in GL(3)$ such that $x_\perp=My_\perp$. $M$ can be decomposed as a dilation by $\frac{|x_\perp|}{|y_\perp|}$ followed by a rotation in the $x_\perp y_\perp$ plane, hence $||M||=\frac{|x_\perp|}{|y_\perp|}$. Let $\tilde{x}_\perp=M^{-1}y_\perp$, so that $y_\perp=M \tilde{x}_\perp$. Now we change integration variable: $q_\perp =M q_\perp '$, $d^3 q_\perp = ||M||^3 \, d^3 q_\perp ' = \frac{|x_\perp|^3}{|y_\perp|^3} \, d^3 q_\perp '$. Since $M$ is invertible and has only the eigenvalue $\frac{|x_\perp|}{|y_\perp|}$, it holds $|A q_\perp '+A z_\perp |^k=||A||^k \cdot |q_\perp '+z_\perp |^k$. After some algebraic manipulations, the second term in \eqref{intform2} reduces to
\begin{equation}\label{secondtermm}
-\frac{3}{\pi} |x_\perp| |y_\perp| \int  d^3 q_\perp   \, \frac{1}{|q_\perp|^2}\cdot  \frac{1}{|y_\perp-q_\perp| \left( |\tilde{x}_\perp-q_\perp|^2  -|y_\perp-q_\perp|^2	\right)},
\end{equation}
which is in the same form of the first integral, with $x_\perp$ replaced by $\tilde{x}_\perp =M^{-1}y_\perp$. Therefore we only need to solve the integral of the first term. Moreover, using conformal invariance, we can set $y_\perp=(1,0,0)^T$ and $x_\perp=(x_1,x_2,0)^T$. Passing to spherical coordinates $(\rho,\theta, \phi)$ this can be rewritten as
\begin{equation}
I(x,y)=-\frac{3}{\pi} |x_\perp| |y_\perp|  \int_0^\infty d \rho \ \int_0^\pi \sin \theta d \theta \, \int_0^{2\pi } d \phi \ \frac{1}{|y_\perp-q_\perp|}\cdot \frac{1}{|x_\perp-q_\perp|^2-|y_\perp-q_\perp|^2}.
\end{equation}
Now the first fraction can be carried out the integration over $\phi$ since $|y_\perp-q_\perp|=\sqrt{\rho^2+1+2\rho\cos\theta}$ is independent of $\phi$, whereas the denominator of the second fraction is of the form $C+ \sin \phi$. Therefore, we can use
$$
\int_0^{2\pi} d \phi \ \frac{1}{C+\sin \phi}=2\pi \frac{\Theta(C^2-1)}{\sqrt{C^2-1}}\cdot \text{sgn}(C),
$$
valid for real $C$, where $\Theta(\cdot )$ is the Heaviside step function.
After straightforward algebraic manipulations, changing variable $\theta \rightarrow t=\cos \theta$, and adding the second piece \eqref{secondtermm}, we find that the integral \eqref{integralll} can be rewritten as
\begin{equation}\label{splitintegral}
I=J(x_1,x_2)+\frac{1}{x_1^2+x_2^2}\cdot  J\left(\frac{x_1}{x_1^2+x_2^2}, \frac{-x_2}{x_1^2+x_2^2} \right),
\end{equation}
with
\begin{equation}
J(x_1,x_2)=-6 \int_0^\infty d\rho \, \int_{-1}^{+1} dt \ \frac{\text{sgn}(C)}{\sqrt{\rho^2+1+2\rho t}} \cdot \frac{\Theta\left((x_1^2+x_2^2-1+2\rho (x_1-1)t)^2-4x_2^2\rho^2 (1-t^2)\right)}{\sqrt{(x_1^2+x_2^2-1+2\rho (x_1-1)t)^2-4x_2^2\rho^2 (1-t^2)}}.
\end{equation}
This integral is very difficult to handle because of the Heaviside step function, which is equivalent to integrating over a very complicated domain. If we restrict to the unit circle $|x_\perp|=x_1^2+x_2^2=1$ (recall that we already set $y_\perp$ to be a unit vector) the integral can be considerably simplified
\begin{equation}
J(x_1,x_2)= -3 \int_0^\infty d\rho \, \int_{-1}^{+1} dt \ \frac{-\text{sgn}(t)}{\rho \cdot\sqrt{\rho^2+1+2\rho t}} \cdot \frac{\Theta\left(2t^2(1-x_1)-x_2^2\right)}{\sqrt{2t^2(1-x_1)-x_2^2}}.
\end{equation}
Now we can rewrite the integral over $t$ as the sum of the integrals from $-1$ to $0$ and from $0$ to $1$ so that the $\text{sgn}(t)$ function disappears, and we can then exchange the order of integration. We also rewrite the Heaviside step function as a boundary in the integral over $t$.
\begin{equation}
J(x_1,x_2)= -\frac{3}{|x_2|} \, \int_{1/ \alpha}^{1} dt \, \frac{1}{\sqrt{\alpha^2 t^2-1}} \int_0^\infty d\rho  \, \frac{1}{\rho }\cdot \left(\frac{1}{ \sqrt{\rho^2+1-2\rho t}}-\frac{1}{ \sqrt{\rho^2+1+2\rho t}}	\right),
\end{equation}
where $\alpha=\frac{\sqrt{2(1-x_1)}}{|x_2|}$. Integrating over $\rho$ yields
\begin{equation}
J(x_1,x_2)= -\frac{3}{|x_2|} \, \int_{1/ \alpha}^{1} dt \, \frac{1}{\sqrt{\alpha^2 t^2-1}} \log \left(\frac{1+t}{1-t} \right).
\end{equation}
Using $x_1^2+x_2^2=1$ and the fact that $J(x_1,x_2)$ does not depend on the sign of $x_2$ we find that the second term in \eqref{splitintegral} is exactly equal to the first, and thus
\begin{equation}
I= -\frac{6}{|x_2|} \, \int_{1/ \alpha}^{1} dt \, \frac{1}{\sqrt{\alpha^2 t^2-1}} \log \left(\frac{1+t}{1-t} \right).
\end{equation}
We can rewrite this in a more convenient way using another change of variable $t=\frac{a^2+s^2}{2a^2 s}$
\begin{equation}
I=-\frac{6}{|x_2|}\int_{\alpha^2-\sqrt{\alpha^4-\alpha^2}}^\alpha ds \, \frac{\log(s^2+2\alpha^2 s + s^2)-\log(-s^2+2\alpha^2 s -\alpha^2)}{\alpha s},
\end{equation}
Finally, solving this integral and passing to lightcone coordinates $z=e^{i \phi}$, $\bar{z}=e^{-i \phi}$ with $0<\phi<\pi$ (note that it is possible to extend the result to the whole circle by simply imposing the symmetry $\phi\rightarrow-\phi$) yields
\begin{equation}
\begin{split}
I\left(z=e^{i\phi},\bar{z}=e^{-i\phi}\right)&=\frac{3}{4} \csc \left(\frac{\phi }{2}\right) \left( 4 \text{Li}_2\left(\frac{2}{\sin \left(\frac{\phi }{2}\right)+1}-1\right)-4 \text{Li}_2\left(\frac{\cos (\phi )+4 \sin \left(\frac{\phi }{2}\right)-3}{\cos (\phi )+1}\right)\right.+\\
& +4 \text{Li}_2\left(-\sec \left(\frac{\phi }{2}\right)-\tan \left(\frac{\phi }{2}\right)\right)-4 \text{Li}_2\left(\sec \left(\frac{\phi }{2}\right)-\tan \left(\frac{\phi }{2}\right)\right)+\\
&+4 \text{Li}_2\left(\tan \left(\frac{\phi }{2}\right)-\sec \left(\frac{\phi }{2}\right)\right)-4 \text{Li}_2\left(\sec \left(\frac{\phi }{2}\right)+\tan \left(\frac{\phi }{2}\right)\right)+\\
&+\pi  \left(\pi -8 i \log \left(\sin \left(\frac{\phi }{4}\right)+\cos \left(\frac{\phi }{4}\right)\right)\right)+8 \log \left(\sin \left(\frac{\phi }{4}\right)+\cos \left(\frac{\phi }{4}\right)\right) \times \\
&\times \left(\log \left(\tan \left(\frac{\phi }{2}\right)\right)-2 \tanh ^{-1}\left(\frac{4 \sin \left(\frac{\phi }{2}\right)+\cos (\phi )-3}{\cos (\phi )+1}\right)\right)+\\
&+8 \log \left(\sec \left(\frac{\phi }{2}\right)\right) \left(\log \left(\frac{2 \sin \left(\frac{\phi }{4}\right)}{\sin \left(\frac{\phi }{2}\right)+\cos \left(\frac{\phi }{2}\right)+1}\right)+\right.\\
&\left.+\left.\tanh ^{-1}\left(\tan \left(\frac{\phi }{2}\right)+\sec \left(\frac{\phi }{2}\right)\right)+\coth ^{-1}\left(\tan \left(\frac{\phi }{2}\right)+\sec \left(\frac{\phi }{2}\right)\right)\right) \right) .
\end{split}
\end{equation}

\label{app:feynman}
\newpage
\bibliographystyle{nb}
\bibliography{references}

\begin{thebibliography}{10}
\ifx\href\asklfhas\newcommand{\href}[2]{#2}\fi
\ifx\arxivref\asklfhas\newcommand{\arxivref}[2]{\href{http://arxiv.org/abs/#1}{#2}}\fi
\ifx\doiref\asklfhas\newcommand{\doiref}[2]{\href{http://dx.doi.org/#1}{#2}}\fi
\raggedright
\small
\parskip 0pt

\bibitem{Rattazzi:2008pe}
R.~Rattazzi, V.~S.~Rychkov, E.~Tonni and A.~Vichi,
\textit{``{Bounding scalar operator dimensions in 4D CFT}''},
\textsf{\doiref{10.1088/1126-6708/2008/12/031}{JHEP~0812,~031~(2008)}},
\texttt{\arxivref{0807.0004}{arxiv:0807.0004}}.

\bibitem{El-Showk:2012cjh}
S.~El-Showk, M.~F.~Paulos, D.~Poland, S.~Rychkov, D.~Simmons-Duffin and
  A.~Vichi,
\textit{``{Solving the 3D Ising Model with the Conformal Bootstrap}''},
\textsf{\doiref{10.1103/PhysRevD.86.025022}{Phys.~Rev.~D~86,~025022~(2012)}},
\texttt{\arxivref{1203.6064}{arxiv:1203.6064}}.

\bibitem{El-Showk:2014dwa}
S.~El-Showk, M.~F.~Paulos, D.~Poland, S.~Rychkov, D.~Simmons-Duffin and
  A.~Vichi,
\textit{``{Solving the 3d Ising Model with the Conformal Bootstrap II.
  c-Minimization and Precise Critical Exponents}''},
\textsf{\doiref{10.1007/s10955-014-1042-7}{J.~Stat.~Phys.~157,~869~(2014)}},
\texttt{\arxivref{1403.4545}{arxiv:1403.4545}}.

\bibitem{Kos:2016ysd}
F.~Kos, D.~Poland, D.~Simmons-Duffin and A.~Vichi,
\textit{``{Precision Islands in the Ising and $O(N)$ Models}''},
\textsf{\doiref{10.1007/JHEP08(2016)036}{JHEP~1608,~036~(2016)}},
\texttt{\arxivref{1603.04436}{arxiv:1603.04436}}.

\bibitem{Poland:2018epd}
D.~Poland, S.~Rychkov and A.~Vichi,
\textit{``{The Conformal Bootstrap: Theory, Numerical Techniques, and
  Applications}''},
\textsf{\doiref{10.1103/RevModPhys.91.015002}{Rev.~Mod.~Phys.~91,~015002~(2019)}},
\texttt{\arxivref{1805.04405}{arxiv:1805.04405}}.

\bibitem{Caron-Huot:2017vep}
S.~Caron-Huot,
\textit{``{Analyticity in Spin in Conformal Theories}''},
\textsf{\doiref{10.1007/JHEP09(2017)078}{JHEP~1709,~078~(2017)}},
\texttt{\arxivref{1703.00278}{arxiv:1703.00278}}.

\bibitem{Simmons-Duffin:2017nub}
D.~Simmons-Duffin, D.~Stanford and E.~Witten,
\textit{``{A spacetime derivation of the Lorentzian OPE inversion formula}''},
\textsf{\doiref{10.1007/JHEP07(2018)085}{JHEP~1807,~085~(2018)}},
\texttt{\arxivref{1711.03816}{arxiv:1711.03816}}.

\bibitem{Carmi:2019cub}
D.~Carmi and S.~Caron-Huot,
\textit{``{A Conformal Dispersion Relation: Correlations from Absorption}''},
\textsf{\doiref{10.1007/JHEP09(2020)009}{JHEP~2009,~009~(2020)}},
\texttt{\arxivref{1910.12123}{arxiv:1910.12123}}.

\bibitem{Alday:2017zzv}
L.~F.~Alday, J.~Henriksson and M.~van~Loon,
\textit{``{Taming the $\epsilon$-expansion with large spin perturbation
  theory}''},
\textsf{\doiref{10.1007/JHEP07(2018)131}{JHEP~1807,~131~(2018)}},
\texttt{\arxivref{1712.02314}{arxiv:1712.02314}}.

\bibitem{Henriksson:2018myn}
J.~Henriksson and M.~Van~Loon,
\textit{``{Critical O(N) model to order $\epsilon^4$ from analytic
  bootstrap}''},
\textsf{\doiref{10.1088/1751-8121/aaf1e2}{J.~Phys.~A~52,~025401~(2019)}},
\texttt{\arxivref{1801.03512}{arxiv:1801.03512}}.

\bibitem{Alday:2019clp}
L.~F.~Alday, J.~Henriksson and M.~van~Loon,
\textit{``{An alternative to diagrams for the critical O(N) model: dimensions
  and structure constants to order 1/N$^{2}$}''},
\textsf{\doiref{10.1007/JHEP01(2020)063}{JHEP~2001,~063~(2020)}},
\texttt{\arxivref{1907.02445}{arxiv:1907.02445}}.

\bibitem{Carmi:2020ekr}
D.~Carmi, J.~Penedones, J.~A.~Silva and A.~Zhiboedov,
\textit{``{Applications of dispersive sum rules: $\epsilon$-expansion and
  holography}''},
\textsf{\doiref{10.21468/SciPostPhys.10.6.145}{SciPost~Phys.~10,~145~(2021)}},
\texttt{\arxivref{2009.13506}{arxiv:2009.13506}}.

\bibitem{Bertucci:2022ptt}
F.~Bertucci, J.~Henriksson and B.~McPeak,
\textit{``{Analytic bootstrap of mixed correlators in the O(n) CFT}''},
\textsf{\doiref{10.1007/JHEP10(2022)104}{JHEP~2210,~104~(2022)}},
\texttt{\arxivref{2205.09132}{arxiv:2205.09132}}.

\bibitem{Liendo:2012hy}
P.~Liendo, L.~Rastelli and B.~C.~van~Rees,
\textit{``{The Bootstrap Program for Boundary CFT$_d$}''},
\textsf{\doiref{10.1007/JHEP07(2013)113}{JHEP~1307,~113~(2013)}},
\texttt{\arxivref{1210.4258}{arxiv:1210.4258}}.

\bibitem{Gliozzi:2015qsa}
F.~Gliozzi, P.~Liendo, M.~Meineri and A.~Rago,
\textit{``{Boundary and Interface CFTs from the Conformal Bootstrap}''},
\textsf{\doiref{10.1007/JHEP05(2015)036}{JHEP~1505,~036~(2015)}},
\texttt{\arxivref{1502.07217}{arxiv:1502.07217}},
[Erratum: JHEP 12, 093 (2021)].

\bibitem{Liendo:2016ymz}
P.~Liendo and C.~Meneghelli,
\textit{``{Bootstrap equations for $ \mathcal{N} $ = 4 SYM with defects}''},
\textsf{\doiref{10.1007/JHEP01(2017)122}{JHEP~1701,~122~(2017)}},
\texttt{\arxivref{1608.05126}{arxiv:1608.05126}}.

\bibitem{deLeeuw:2017dkd}
M.~de~Leeuw, A.~C.~Ipsen, C.~Kristjansen, K.~E.~Vardinghus and M.~Wilhelm,
\textit{``{Two-point functions in AdS/dCFT and the boundary conformal bootstrap
  equations}''},
\textsf{\doiref{10.1007/JHEP08(2017)020}{JHEP~1708,~020~(2017)}},
\texttt{\arxivref{1705.03898}{arxiv:1705.03898}}.

\bibitem{Rastelli:2017ecj}
L.~Rastelli and X.~Zhou,
\textit{``{The Mellin Formalism for Boundary CFT$_d$}''},
\textsf{\doiref{10.1007/JHEP10(2017)146}{JHEP~1710,~146~(2017)}},
\texttt{\arxivref{1705.05362}{arxiv:1705.05362}}.

\bibitem{defectABJM}
L.~Bianchi, L.~Griguolo, M.~Preti and D.~Seminara,
\textit{``{Wilson lines as superconformal defects in ABJM theory: a formula for
  the emitted radiation}''},
\textsf{\doiref{10.1007/JHEP10(2017)050}{JHEP~1710,~050~(2017)}},
\texttt{\arxivref{1706.06590}{arxiv:1706.06590}}.

\bibitem{Drukker:2017dgn}
N.~Drukker, I.~Shamir and C.~Vergu,
\textit{``{Defect multiplets of $ \mathcal{N}=1 $ supersymmetry in 4d}''},
\textsf{\doiref{10.1007/JHEP01(2018)034}{JHEP~1801,~034~(2018)}},
\texttt{\arxivref{1711.03455}{arxiv:1711.03455}}.

\bibitem{Giombi:2018qox}
S.~Giombi and S.~Komatsu,
\textit{``{Exact Correlators on the Wilson Loop in $\mathcal{N}=4$ SYM:
  Localization, Defect CFT, and Integrability}''},
\textsf{\doiref{10.1007/JHEP05(2018)109}{JHEP~1805,~109~(2018)}},
\texttt{\arxivref{1802.05201}{arxiv:1802.05201}},
[Erratum: JHEP 11, 123 (2018)].

\bibitem{Bianchi:2018scb}
L.~Bianchi, M.~Preti and E.~Vescovi,
\textit{``{Exact Bremsstrahlung functions in ABJM theory}''},
\textsf{\doiref{10.1007/JHEP07(2018)060}{JHEP~1807,~060~(2018)}},
\texttt{\arxivref{1802.07726}{arxiv:1802.07726}}.

\bibitem{Bianchi:2018zpb}
L.~Bianchi, M.~Lemos and M.~Meineri,
\textit{``{Line Defects and Radiation in $\mathcal{N}=2$ Conformal
  Theories}''},
\textsf{Phys.~Rev.~Lett.~121,~141601~(2018)},
\texttt{\arxivref{1805.04111}{arxiv:1805.04111}}.

\bibitem{Carlo}
P.~Liendo, C.~Meneghelli and V.~Mitev,
\textit{``Bootstrapping the half-BPS line defect''},
\textsf{\doiref{10.1007/JHEP10(2018)077}{JHEP~1810,~077~(2018)}},
\texttt{\arxivref{1806.01862}{arxiv:1806.01862}}.

\bibitem{Bissi:2018mcq}
A.~Bissi, T.~Hansen and A.~S\"oderberg,
\textit{``{Analytic Bootstrap for Boundary CFT}''},
\textsf{\doiref{10.1007/JHEP01(2019)010}{JHEP~1901,~010~(2019)}},
\texttt{\arxivref{1808.08155}{arxiv:1808.08155}}.

\bibitem{Kaviraj:2018tfd}
A.~Kaviraj and M.~F.~Paulos,
\textit{``{The Functional Bootstrap for Boundary CFT}''},
\textsf{\doiref{10.1007/JHEP04(2020)135}{JHEP~2004,~135~(2020)}},
\texttt{\arxivref{1812.04034}{arxiv:1812.04034}}.

\bibitem{Mazac:2018biw}
D.~Maz\'a\v{c}, L.~Rastelli and X.~Zhou,
\textit{``{An analytic approach to BCFT$_{d}$}''},
\textsf{\doiref{10.1007/JHEP12(2019)004}{JHEP~1912,~004~(2019)}},
\texttt{\arxivref{1812.09314}{arxiv:1812.09314}}.

\bibitem{DiPietro:2019hqe}
L.~Di~Pietro, D.~Gaiotto, E.~Lauria and J.~Wu,
\textit{``{3d Abelian Gauge Theories at the Boundary}''},
\textsf{\doiref{10.1007/JHEP05(2019)091}{JHEP~1905,~091~(2019)}},
\texttt{\arxivref{1902.09567}{arxiv:1902.09567}}.

\bibitem{Gimenez-Grau:2019hez}
A.~Gimenez-Grau and P.~Liendo,
\textit{``{Bootstrapping line defects in $\mathcal{N}=2$ theories}''},
\textsf{\doiref{10.1007/JHEP03(2020)121}{JHEP~2003,~121~(2020)}},
\texttt{\arxivref{1907.04345}{arxiv:1907.04345}}.

\bibitem{Bianchi:2019dlw}
L.~Bianchi, M.~Billo, F.~Galvagno and A.~Lerda,
\textit{``{Emitted Radiation and Geometry}''},
\textsf{\doiref{10.1007/JHEP01(2020)075}{JHEP~2001,~075~(2020)}},
\texttt{\arxivref{1910.06332}{arxiv:1910.06332}}.

\bibitem{Bianchi:2019sxz}
L.~Bianchi and M.~Lemos,
\textit{``{Superconformal surfaces in four dimensions}''},
\textsf{\doiref{10.1007/JHEP06(2020)056}{JHEP~2006,~056~(2020)}},
\texttt{\arxivref{1911.05082}{arxiv:1911.05082}}.

\bibitem{Giombi:2019enr}
S.~Giombi and H.~Khanchandani,
\textit{``{$O(N)$ models with boundary interactions and their long range
  generalizations}''},
\textsf{\doiref{10.1007/JHEP08(2020)010}{JHEP~2008,~010~(2020)}},
\texttt{\arxivref{1912.08169}{arxiv:1912.08169}}.

\bibitem{Wang:2020seq}
Y.~Wang,
\textit{``{Taming defects in $ \mathcal{N} $ = 4 super-Yang-Mills}''},
\textsf{\doiref{10.1007/JHEP08(2020)021}{JHEP~2008,~021~(2020)}},
\texttt{\arxivref{2003.11016}{arxiv:2003.11016}}.

\bibitem{Bianchi:2020hsz}
L.~Bianchi, G.~Bliard, V.~Forini, L.~Griguolo and D.~Seminara,
\textit{``{Analytic bootstrap and Witten diagrams for the ABJM Wilson line as
  defect CFT$_{1}$}''},
\textsf{\doiref{10.1007/JHEP08(2020)143}{JHEP~2008,~143~(2020)}},
\texttt{\arxivref{2004.07849}{arxiv:2004.07849}}.

\bibitem{Ashok:2020ekv}
S.~K.~Ashok, M.~Billo, M.~Frau, A.~Lerda and S.~Mahato,
\textit{``{Surface defects from fractional branes. Part I}''},
\textsf{\doiref{10.1007/JHEP07(2020)051}{JHEP~2007,~051~(2020)}},
\texttt{\arxivref{2005.02050}{arxiv:2005.02050}}.

\bibitem{Lauria:2020emq}
E.~Lauria, P.~Liendo, B.~C.~Van~Rees and X.~Zhao,
\textit{``{Line and surface defects for the free scalar field}''},
\textsf{\doiref{10.1007/JHEP01(2021)060}{JHEP~2101,~060~(2021)}},
\texttt{\arxivref{2005.02413}{arxiv:2005.02413}}.

\bibitem{Behan:2020nsf}
C.~Behan, L.~Di~Pietro, E.~Lauria and B.~C.~Van~Rees,
\textit{``{Bootstrapping boundary-localized interactions}''},
\textsf{\doiref{10.1007/JHEP12(2020)182}{JHEP~2012,~182~(2020)}},
\texttt{\arxivref{2009.03336}{arxiv:2009.03336}}.

\bibitem{Agmon:2020pde}
N.~B.~Agmon and Y.~Wang,
\textit{``{Classifying Superconformal Defects in Diverse Dimensions Part I:
  Superconformal Lines}''},
\texttt{\arxivref{2009.06650}{arxiv:2009.06650}}.

\bibitem{Drukker:2020atp}
N.~Drukker, M.~Probst and M.~Tr\'epanier,
\textit{``{Defect CFT techniques in the 6d $\mathcal{N} = (2,0)$ theory}''},
\textsf{\doiref{10.1007/JHEP03(2021)261}{JHEP~2103,~261~(2021)}},
\texttt{\arxivref{2009.10732}{arxiv:2009.10732}}.

\bibitem{Herzog:2020bqw}
C.~P.~Herzog and A.~Shrestha,
\textit{``{Two point functions in defect CFTs}''},
\textsf{\doiref{10.1007/JHEP04(2021)226}{JHEP~2104,~226~(2021)}},
\texttt{\arxivref{2010.04995}{arxiv:2010.04995}}.

\bibitem{Barrat:2020vch}
J.~Barrat, P.~Liendo and J.~Plefka,
\textit{``{Two-point correlator of chiral primary operators with a Wilson line
  defect in $ \mathcal{N} $ = 4 SYM}''},
\textsf{\doiref{10.1007/JHEP05(2021)195}{JHEP~2105,~195~(2021)}},
\texttt{\arxivref{2011.04678}{arxiv:2011.04678}}.

\bibitem{Dey:2020jlc}
P.~Dey and A.~S\"oderberg,
\textit{``{On analytic bootstrap for interface and boundary CFT}''},
\textsf{\doiref{10.1007/JHEP07(2021)013}{JHEP~2107,~013~(2021)}},
\texttt{\arxivref{2012.11344}{arxiv:2012.11344}}.

\bibitem{Giombi:2021uae}
S.~Giombi, E.~Helfenberger, Z.~Ji and H.~Khanchandani,
\textit{``{Monodromy defects from hyperbolic space}''},
\textsf{\doiref{10.1007/JHEP02(2022)041}{JHEP~2202,~041~(2022)}},
\texttt{\arxivref{2102.11815}{arxiv:2102.11815}}.

\bibitem{Ferrero:2021bsb}
P.~Ferrero and C.~Meneghelli,
\textit{``{Bootstrapping the half-BPS line defect CFT in $\mathcal{N}=4$ SYM at
  strong coupling}''},
\texttt{\arxivref{2103.10440}{arxiv:2103.10440}}.

\bibitem{Bianchi:2021snj}
L.~Bianchi, A.~Chalabi, V.~Proch\'azka, B.~Robinson and J.~Sisti,
\textit{``{Monodromy defects in free field theories}''},
\textsf{\doiref{10.1007/JHEP08(2021)013}{JHEP~2108,~013~(2021)}},
\texttt{\arxivref{2104.01220}{arxiv:2104.01220}}.

\bibitem{Bianchi:2021piu}
L.~Bianchi, G.~Bliard, V.~Forini and G.~Peveri,
\textit{``{Mellin amplitudes for 1d CFT}''},
\textsf{\doiref{10.1007/JHEP10(2021)095}{JHEP~2110,~095~(2021)}},
\texttt{\arxivref{2106.00689}{arxiv:2106.00689}}.

\bibitem{Cavaglia:2021bnz}
A.~Cavagli\`a, N.~Gromov, J.~Julius and M.~Preti,
\textit{``{Integrability and Conformal Bootstrap: One Dimensional Defect
  CFT}''},
\texttt{\arxivref{2107.08510}{arxiv:2107.08510}}.

\bibitem{Gimenez-Grau:2021wiv}
A.~Gimenez-Grau and P.~Liendo,
\textit{``{Bootstrapping Monodromy Defects in the Wess-Zumino Model}''},
\texttt{\arxivref{2108.05107}{arxiv:2108.05107}}.

\bibitem{Barrat:2021yvp}
J.~Barrat, A.~Gimenez-Grau and P.~Liendo,
\textit{``{Bootstrapping holographic defect correlators in $ \mathcal{N} $ = 4
  super Yang-Mills}''},
\textsf{\doiref{10.1007/JHEP04(2022)093}{JHEP~2204,~093~(2022)}},
\texttt{\arxivref{2108.13432}{arxiv:2108.13432}}.

\bibitem{Padayasi:2021sik}
J.~Padayasi, A.~Krishnan, M.~A.~Metlitski, I.~A.~Gruzberg and M.~Meineri,
\textit{``{The extraordinary boundary transition in the 3d O(N) model via
  conformal bootstrap}''},
\texttt{\arxivref{2111.03071}{arxiv:2111.03071}}.

\bibitem{Behan:2021tcn}
C.~Behan, L.~Di~Pietro, E.~Lauria and B.~C.~van~Rees,
\textit{``{Bootstrapping boundary-localized interactions II. Minimal models at
  the boundary}''},
\textsf{\doiref{10.1007/JHEP03(2022)146}{JHEP~2203,~146~(2022)}},
\texttt{\arxivref{2111.04747}{arxiv:2111.04747}}.

\bibitem{Collier:2021ngi}
S.~Collier, D.~Mazac and Y.~Wang,
\textit{``{Bootstrapping Boundaries and Branes}''},
\texttt{\arxivref{2112.00750}{arxiv:2112.00750}}.

\bibitem{Herzog:2022jqv}
C.~P.~Herzog and A.~Shrestha,
\textit{``{Conformal Surface Defects in Maxwell Theory are Trivial}''},
\texttt{\arxivref{2202.09180}{arxiv:2202.09180}}.

\bibitem{Cavaglia:2022qpg}
A.~Cavagli\`a, N.~Gromov, J.~Julius and M.~Preti,
\textit{``{Bootstrability in Defect CFT: Integrated Correlators and Sharper
  Bounds}''},
\texttt{\arxivref{2203.09556}{arxiv:2203.09556}}.

\bibitem{Chalabi:2022qit}
A.~Chalabi, C.~P.~Herzog, K.~Ray, B.~Robinson, J.~Sisti and A.~Stergiou,
\textit{``{Boundaries in Free Higher Derivative Conformal Field Theories}''},
\texttt{\arxivref{2211.14335}{arxiv:2211.14335}}.

\bibitem{Billo:2016cpy}
M.~Billo, V.~Goncalves, E.~Lauria and M.~Meineri,
\textit{``{Defects in conformal field theory}''},
\textsf{\doiref{10.1007/JHEP04(2016)091}{JHEP~1604,~091~(2016)}},
\texttt{\arxivref{1601.02883}{arxiv:1601.02883}}.

\bibitem{Lemos:2017vnx}
M.~Lemos, P.~Liendo, M.~Meineri and S.~Sarkar,
\textit{``{Universality at large transverse spin in defect CFT}''},
\textsf{\doiref{10.1007/JHEP09(2018)091}{JHEP~1809,~091~(2018)}},
\texttt{\arxivref{1712.08185}{arxiv:1712.08185}}.

\bibitem{Liendo:2019jpu}
P.~Liendo, Y.~Linke and V.~Schomerus,
\textit{``{A Lorentzian inversion formula for defect CFT}''},
\textsf{\doiref{10.1007/JHEP08(2020)163}{JHEP~2008,~163~(2020)}},
\texttt{\arxivref{1903.05222}{arxiv:1903.05222}}.

\bibitem{Barrat:2022psm}
J.~Barrat, A.~Gimenez-Grau and P.~Liendo,
\textit{``{A dispersion relation for defect CFT}''},
\texttt{\arxivref{2205.09765}{arxiv:2205.09765}}.

\bibitem{Bianchi:2022ppi}
L.~Bianchi and D.~Bonomi,
\textit{``{Conformal dispersion relations for defects and boundaries}''},
\texttt{\arxivref{2205.09775}{arxiv:2205.09775}}.

\bibitem{Allaismagnetic}
A.~Allais,
\textit{``Magnetic defect line in a critical Ising bath''},
\href{https://arxiv.org/abs/1412.3449}{\texttt{https://arxiv.org/abs/1412.3449}}.

\bibitem{ParisenToldin2017}
F.~P.~Toldin, F.~F.~Assaad and S.~Wessel,
\textit{``Critical behavior in the presence of an order-parameter pinning
  field''},
\textsf{\doiref{10.1103/physrevb.95.014401}{Physical~Review~B~95,~~(2017)}}.

\bibitem{Ebadi:2020ldi}
S.~Ebadi et~al.,
\textit{``{Quantum phases of matter on a 256-atom programmable quantum
  simulator}''},
\textsf{\doiref{10.1038/s41586-021-03582-4}{Nature~595,~227~(2021)}},
\texttt{\arxivref{2012.12281}{arxiv:2012.12281}}.

\bibitem{LAW2001159}
B.~M.~Law,
\textit{``Wetting, adsorption and surface critical phenomena''},
\textsf{Progress~in~Surface~Science~66,~159~(2001)}.

\bibitem{Fisher2003}
M.~Fisher and P.~de~Gennes,
\textit{``{ Phenomenes aux parois dans un melange binaire critique}''},
\textsf{\doiref{10.1142/9789812564849 0025}{World~Scientific~Simple Views on
  Condensed Matter,~237–241~(2003)}}.

\bibitem{PhysRevLett.84.2180}
A.~Hanke,
\textit{``Critical Adsorption on Defects in Ising Magnets and Binary Alloys''},
\textsf{\doiref{10.1103/PhysRevLett.84.2180}{Phys.~Rev.~Lett.~84,~2180~(2000)}},
\href{https://link.aps.org/doi/10.1103/PhysRevLett.84.2180}{\texttt{https://link.aps.org/doi/10.1103/PhysRevLett.84.2180}}.

\bibitem{Allais:2014fqa}
A.~Allais and S.~Sachdev,
\textit{``{Spectral function of a localized fermion coupled to the
  Wilson-Fisher conformal field theory}''},
\textsf{\doiref{10.1103/PhysRevB.90.035131}{Phys.~Rev.~B~90,~035131~(2014)}},
\texttt{\arxivref{1406.3022}{arxiv:1406.3022}}.

\bibitem{Cuomo:2021kfm}
G.~Cuomo, Z.~Komargodski and M.~Mezei,
\textit{``{Localized magnetic field in the O(N) model}''},
\textsf{\doiref{10.1007/JHEP02(2022)134}{JHEP~2202,~134~(2022)}},
\texttt{\arxivref{2112.10634}{arxiv:2112.10634}}.

\bibitem{Gimenez-Grau:2022czc}
A.~Gimenez-Grau, E.~Lauria, P.~Liendo and P.~van~Vliet,
\textit{``{Bootstrapping line defects with O(2) global symmetry}''},
\textsf{\doiref{10.1007/JHEP11(2022)018}{JHEP~2211,~018~(2022)}},
\texttt{\arxivref{2208.11715}{arxiv:2208.11715}}.

\bibitem{sengupta97}
A.~M.~Sengupta,
\textit{``Spin in a Fluctuating Field: The Bose (+Fermi) Kondo models''},
\href{https://arxiv.org/abs/cond-mat/9707316}{\texttt{https://arxiv.org/abs/cond-mat/9707316}}.

\bibitem{Vojta_2000}
M.~Vojta, C.~Buragohain and S.~Sachdev,
\textit{``Quantum impurity dynamics in two-dimensional antiferromagnets and
  superconductors''},
\textsf{\doiref{10.1103/physrevb.61.15152}{Physical~Review~B~61,~15152~(2000)}}.

\bibitem{Sachdev_2000}
S.~Sachdev, C.~Buragohain and M.~Vojta,
\textit{``Quantum Impurity in a Nearly Critical Two Dimensional
  Antiferromagnet''},
\href{https://arxiv.org/abs/cond-mat/0004156}{\texttt{https://arxiv.org/abs/cond-mat/0004156}}.

\bibitem{Sachdev:2001ky}
S.~Sachdev,
\textit{``{Static hole in a critical antiferromagnet: Field theoretic
  renormalization group}''},
\textsf{\doiref{10.1016/S0921-4534(01)00198-8}{Physica~C~357,~78~(2001)}},
\texttt{\arxivref{cond-mat/0011233}{cond-mat/0011233}}.

\bibitem{Sachdev:2003yk}
S.~Sachdev and M.~Vojta,
\textit{``{Quantum impurity in an antiferromagnet: Nonlinear sigma model
  theory}''},
\textsf{\doiref{10.1103/PhysRevB.68.064419}{Phys.~Rev.~B~68,~064419~(2003)}},
\texttt{\arxivref{cond-mat/0303001}{cond-mat/0303001}}.

\bibitem{Florens_2006}
S.~Florens, L.~Fritz and M.~Vojta,
\textit{``Kondo Effect in Bosonic Spin Liquids''},
\textsf{\doiref{10.1103/physrevlett.96.036601}{Physical~Review~Letters~96,~~(2006)}}.

\bibitem{Florens_2007}
S.~Florens, L.~Fritz and M.~Vojta,
\textit{``Boundary quantum criticality in models of magnetic impurities coupled
  to bosonic baths''},
\textsf{\doiref{10.1103/physrevb.75.224420}{Physical~Review~B~75,~~(2007)}}.

\bibitem{Liu_2021}
S.~Liu, H.~Shapourian, A.~Vishwanath and M.~A.~Metlitski,
\textit{``Magnetic impurities at quantum critical points: Large N expansion and
  connections to symmetry protected topological states''},
\textsf{\doiref{10.1103/physrevb.104.104201}{Physical~Review~B~104,~~(2021)}}.

\bibitem{Gaiotto:2013nva}
D.~Gaiotto, D.~Mazac and M.~F.~Paulos,
\textit{``{Bootstrapping the 3d Ising twist defect}''},
\textsf{\doiref{10.1007/JHEP03(2014)100}{JHEP~1403,~100~(2014)}},
\texttt{\arxivref{1310.5078}{arxiv:1310.5078}}.

\bibitem{Billo:2013jda}
M.~Billò, M.~Caselle, D.~Gaiotto, F.~Gliozzi, M.~Meineri and R.~Pellegrini,
\textit{``{Line defects in the 3d Ising model}''},
\textsf{\doiref{10.1007/JHEP07(2013)055}{JHEP~1307,~055~(2013)}},
\texttt{\arxivref{1304.4110}{arxiv:1304.4110}}.

\bibitem{Giombi:2022vnz}
S.~Giombi, E.~Helfenberger and H.~Khanchandani,
\textit{``{Line Defects in Fermionic CFTs}''},
\texttt{\arxivref{2211.11073}{arxiv:2211.11073}}.

\bibitem{toappear}
A.~Gimenez-Grau,
\textit{``{Probing magnetic line defects with two-point functions}''}.

\bibitem{Wilson:1971dc}
K.~G.~Wilson and M.~E.~Fisher,
\textit{``{Critical exponents in 3.99 dimensions}''},
\textsf{\doiref{10.1103/PhysRevLett.28.240}{Phys.~Rev.~Lett.~28,~240~(1972)}}.

\bibitem{Henriksson:2022rnm}
J.~Henriksson,
\textit{``{The critical O(N) CFT: Methods and conformal data}''},
\texttt{\arxivref{2201.09520}{arxiv:2201.09520}}.

\bibitem{Kapustin:2005py}
A.~Kapustin,
\textit{``{Wilson-'t Hooft operators in four-dimensional gauge theories and
  S-duality}''},
\textsf{\doiref{10.1103/PhysRevD.74.025005}{Phys.~Rev.~D~74,~025005~(2006)}},
\texttt{\arxivref{hep-th/0501015}{hep-th/0501015}}.

\bibitem{Isachenkov:2018pef}
M.~Isachenkov, P.~Liendo, Y.~Linke and V.~Schomerus,
\textit{``{Calogero-Sutherland Approach to Defect Blocks}''},
\textsf{\doiref{10.1007/JHEP10(2018)204}{JHEP~1810,~204~(2018)}},
\texttt{\arxivref{1806.09703}{arxiv:1806.09703}}.

\bibitem{Giombi:2016hkj}
S.~Giombi and V.~Kirilin,
\textit{``{Anomalous dimensions in CFT with weakly broken higher spin
  symmetry}''},
\textsf{\doiref{10.1007/JHEP11(2016)068}{JHEP~1611,~068~(2016)}},
\texttt{\arxivref{1601.01310}{arxiv:1601.01310}}.

\bibitem{Alday:2017vkk}
L.~F.~Alday and S.~Caron-Huot,
\textit{``{Gravitational S-matrix from CFT dispersion relations}''},
\textsf{\doiref{10.1007/JHEP12(2018)017}{JHEP~1812,~017~(2018)}},
\texttt{\arxivref{1711.02031}{arxiv:1711.02031}}.

\bibitem{KdFconvergence}
H.~M.~Srivastava and M.~C.~Daoust,
\textit{``A Note on the Convergence of KAMPÉ DE FÉRIET's Double
  Hypergeometric Series''},
\textsf{Mathematische~Nachrichten~53,~151~(1972)}.

\end{thebibliography}

\end{document}